# Fast Accurate Point of Care COVID-19 Pandemic Diagnosis Enabled Through Advanced Lab-on-a-Chip Optical Biosensors: Opportunities and Challenges

Aref Asghari[1, †], Chao Wang[1, †], Kyoung Min Yoo[1, †], Hamed Dalir[1, 2, *] and Ray T. Chen[1, 2, *]

[1]*Department of Electrical and Computer Engineering, The University of Texas at Austin, Austin, TX 78758, USA.*
[2]*Omega Optics, Inc. 8500 Shoal Creek Blvd., Austin, Texas 78757, USA*

[*] Corresponding Authors Emails: hamed.dalir@omegaoptics.com , chenrt@austin.utexas.edu
[†] These authors have equal contributions to this work.

## Abstract

**The sudden rise of severe acute respiratory syndrome coronavirus 2 (SARS-CoV-2) pandemic early 2020 throughout the world has called into drastic action measures to do instant detection and reduce the spread rate. The common diagnostics testing methods has been only partially effective in satisfying the booming demand for fast detection methods to contain the further spread. However, the point-of-risk accurate diagnosis of this new emerging viral infection is paramount as simultaneous normal working operation and dealing with symptoms of SARS-CoV-2 can become the norm for years to come. Sensitive cost-effective biosensor with mass production capability is crucial throughout the world until a universal vaccination become available. Optical label-free biosensors can provide a non-invasive, extremely sensitive rapid detection technique up to ~1 fM ($10^{-15}$) concentration along with few minutes sensing. These biosensors can be manufactured on a mass-scale (billions) to detect the COVID-19 viral load in nasal, saliva, urinal, and serological samples even if the infected person is asymptotic. Methods investigated here are the most advanced available platforms for biosensing optical devices resulted from the integration of state-of-the-art designs and materials. These approaches are including but not limited to integrated optical devices, plasmonic resonance and also emerging nanomaterial biosensors. The lab-on-a-chip platforms examined here are suitable not only for SARS-CoV-2 spike protein detection but also other contagious virions such as influenza, and middle east respiratory syndrome (MERS).**

## I. INTRODUCTION

A. Importance of highly sensitive point of care detection

The Corona virus disease 2019 (COVID-19) pandemic and its rapid growth rate has driven an unprecedented worldwide demand for measures to mitigate its fast spread rate[1–3]. Adding mandatory large-scale policies like social distancing and extreme antiviral disinfection measures and protocols in coping with infected patients, it becomes paramount to have detection and curing system developed in a catastrophic crisis state of which very few mankind alive have experienced before. It is a commonly held view that pressures of war have stimulated advances in engineering, science, and medicine. Therefore, the new invisible battle against SARS-CoV-2 virus infection can stimulate major breakthroughs in the development of diagnosis and treatment systems. SARS-CoV-2 highly contagious infection is hard to detect as patients can be present with clinically inapparent symptoms including fever, cough, or shortness of breath[4]. The worldwide morbidity and mortality of SARS-CoV-2 plus no available vaccine or guaranteed treatments on the horizon as of mid-2020 bolds the necessity for researchers to probe various medical interventions. Immediate cost-effective point-of-risk measures like identification, diagnosis and isolation of the infected individual is still regarded as the single best viable solution to slow down this pneumonia pandemic.

B. Structure of the SARS-CoV-2

SARS-CoV is an enveloped, single stranded RNA virus, that exists in humans and animals, and is mainly transmitted through aerosols and nearby interpersonal contacts [5,6]. Once the virus enters the body, it sticks to primary target cells which provides plenty of virus receptors, the angiotensin-converting enzyme (ACE2) [5,7]. Its genome RNA infusion into the cell results in the formation of protein building blocks consist of spike, envelope, membrane, nucleocapsid, and proteins[8–10]. Thus, human SARS-CoVs relies heavily on ACE2 for infusion into the target cell for which S glycoprotein trimeric spikes on the surface mediates the entrance into the host cell[11]. The S glycoprotein of SARS-CoV is therefore the main target for neutralizing antibodies (nAbs) [12]. Similar SARS-CoV and SARS-CoV-2 amino acid identity in their S proteins makes them prone to have analogous Immunogenic surfaces on these antigens[13]. Coronaviruses demonstrate a complex pattern for receptor recognition[7,14,15] (Figure 1.c). The attempts to block the infusion of virus has been carried out through targeting mainly spike protein of SARS-CoV-2 and the receptor binding domain (RBD). Antibodies developed specifically for these regions can expand the potency and power and chance of success against the infusion of SARS-CoV-2 in the host cell. Once the virus enters the body through the cells, it replicates and virions are then set free to infect new target cells[16,17]. SARS Infectious viral particles can be found in respiratory secretions, urine and sweat. SARS-CoV infection harms lung tissues resulting in pneumonia with rapid respiratory deterioration and failure and in almost %5 of cases, death[18,19]. Development of effective vaccine against SARS-CoV-2 can be effectively applied through S protein and especially the receptor binding domain (RBD) as they induce highly potent neutralizing antibody to block virus binding and its membrane infusion or forming immunity protective layer against viral infection[5].

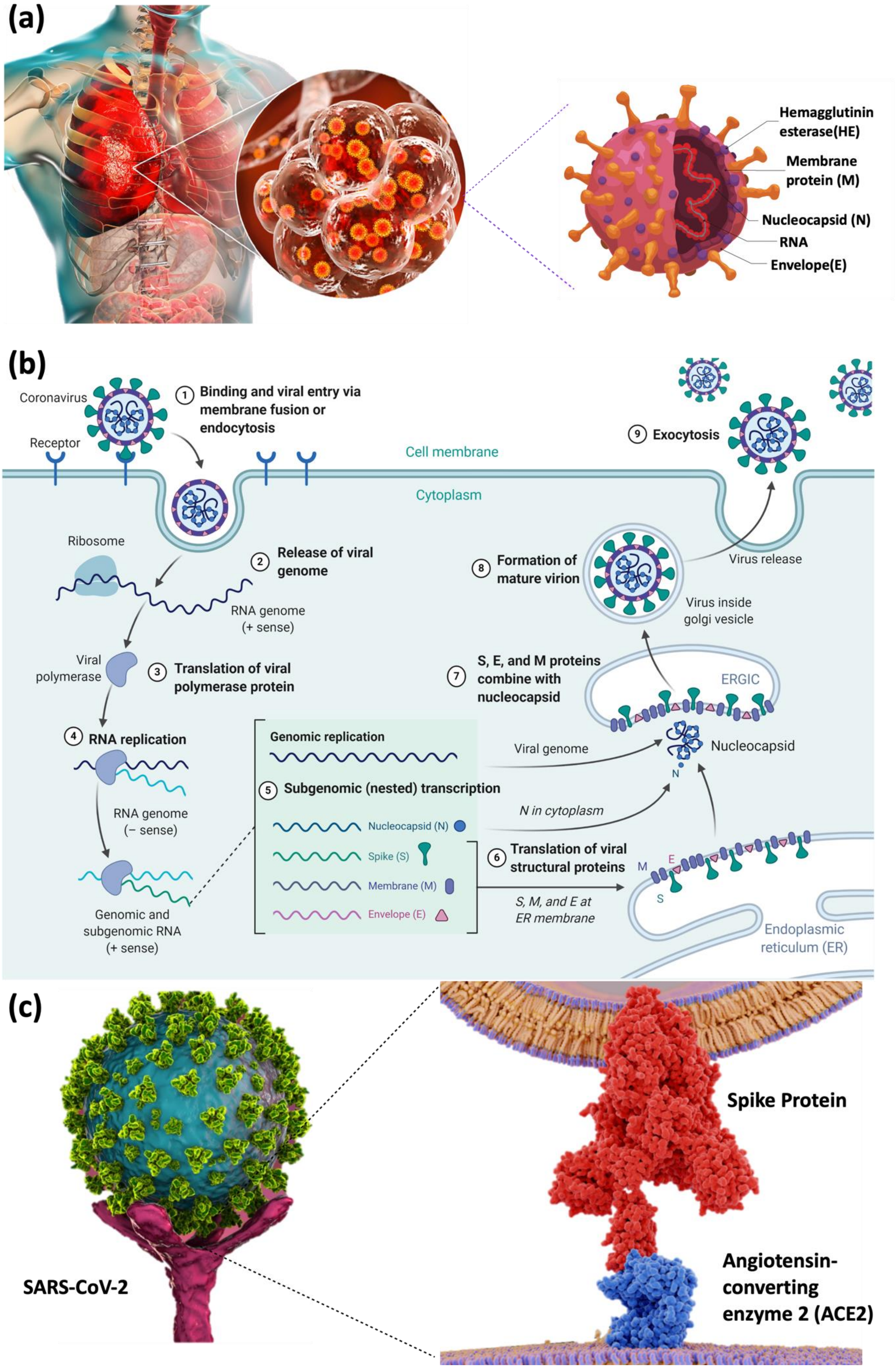
(a)
Hemagglutinin esterase(HE)
Membrane protein (M)
Nucleocapsid (N)
RNA
Envelope(E)
(b)
Coronavirus
Receptor
1 Binding and viral entry via membrane fusion or endocytosis
Cell membrane
Cytoplasm
9 Exocytosis
Ribosome
2 Release of viral genome
RNA genome (+ sense)
8 Formation of mature virion
Virus release
Virus inside golgi vesicle
Viral polymerase
3 Translation of viral polymerase protein
7 S, E, and M proteins combine with nucleocapsid
ERGIC
4 RNA replication
Genomic replication
Viral genome
Nucleocapsid
N
5 Subgenomic (nested) transcription
RNA genome (− sense)
N in cytoplasm
Nucleocapsid (N)
Spike (S)
Membrane (M)
Envelope (E)
6 Translation of viral structural proteins
M
E
S
S, M, and E at ER membrane
Genomic and subgenomic RNA (+ sense)
Endoplasmic reticulum (ER)
(c)
Spike Protein
SARS-CoV-2
Angiotensin-converting enzyme 2 (ACE2)

**Fig 1. (a)** Colorized schematic picture depicting a heavily infected lung by SARS-COV-2 virus molecules. SARS-COV-2 molecular structure is illustrated in detail as well: the RNA and membrane protein are significant as they provide great affinity to bio receptors functionalized on the surface of biosensor. (b) (A) Life cycle of pathogenic human SARS-CoVs. The virus enters the target cell through respective cellular receptor angiotensin-converting enzyme 2 (ACE2) on the membranes of host cells. Viral genomic RNA is then unveiled in the cytoplasm and translated into viral polymerase proteins. Viral RNA and nucleocapsid (N) structural protein are replicated and transcribed in the cytoplasm to form a mature virion, then released from host cells[12]. (c) The virus enters the target cell by first binding its S glycoproteins to the respective cellular receptor angiotensin-converting enzyme 2 (ACE2) on the membranes of host cells which mediate virus-cell membrane fusion and viral entry, (left) Schematic of SARS-CoV-2 virus binding to ACE-2 receptors on a human cell (right) Schematic shows the coronavirus spike protein (red) mediates the virus entry into host cells. It binds to the angiotensin converting enzyme 2 (blue) and fuses viral and host membranes.

## C. Detection mechanisms for SARS-CoV-2 viral infection

The most standard procedure for identifying pathogens like SARS-CoV-2 relies on real-time reverse transcription- polymerase chain reaction (RT-PCR) in which the virus ribonucleic acid (RNA) molecules (Fig. 1) go through a time consuming labeling procedure known as reverse transcription (RT) [20]. For patients who present late with a viral load below the detection limit of RT-PCR assays, serological diagnosis with less sensitivity is also used. The aforementioned process and similar clinical diagnosis requires advanced laboratories, equipment and expertise hard to be found in under-developed remote areas that are more prone to the outbreak[20–24].

The test is done for a qualitative analysis of nucleic acid from the SARS-CoV-2 gathered from people who meet SARS-CoV-2 virus clinical infection signs and symptoms (Fig 1.b). It has also been reported that SARS-CoV-2 can be clinically detected from saliva, blood, and urine samples in 30 to 45 minutes. Considering the high cost and time-consuming nature of clinical diagnosis procedure like RT-PCR (minimum 3 hrs[25]) and test kit requiring samples generated by urine, saliva, blood and nasopharyngeal swabs (minimum 30 minutes[9–11]), the need to develop a fast-accurate detection method for SARS-CoV-2 is better recognized. Biosensors can provide the next best alternative reliable solution to clinical diagnosis with much faster real-time detection without compromising sensitivity and accuracy[29,30]. Optical biosensors are particularly likely to become the future COVID-19 diagnostic tools[31–35]. By exploiting the strong light-matter interactions, one can create an ultra-sensitive label-free real-time detection platform for novel SARS-COV-2. Primarily, an optical biosensor translates the capture of the target analyte in a measurable alteration of a light property, such as refractive index (RI), intensity or resonance shift, through different methods such as resonators and interferometers (Fig. 2).

However, due to time-sensitivity and shocking nature of the COVID-19 pandemic, early efforts for detection have been mainly based on how existing systems can be integrated together competently to outperform the existing processes in respect to sensitivity and time-consumption. It has been demonstrated by Chen et al. and Alam et al. [36,37] that deep learning methods can be efficiently used to aid radiologists in real-time accurate diagnosis of COVID-19 infection from computed tomography (CT) images. Artificial intelligence (AI) technology have been implemented to thoroughly process a large group of patient's characteristics by four different algorithms to come up with the 18 diagnostic factors relating to COVID-19[38]. Zhang et al. reported the use of graphene field effect transistors (FETs) in combination with selective antibodies to develop coronavirus immunosensors[39]. Combination of plasmonic photothermal effect and localized surface plasmon resonance sensing transduction has also been proposed as promising for COVID-19 diagnosis[29].

D. Principle of lab-on-a-chip optical biosensors

To spot and monitor the real-time binding of small numbers of biomolecules such as proteins, biosensors with ultra-high sensitivity are required.

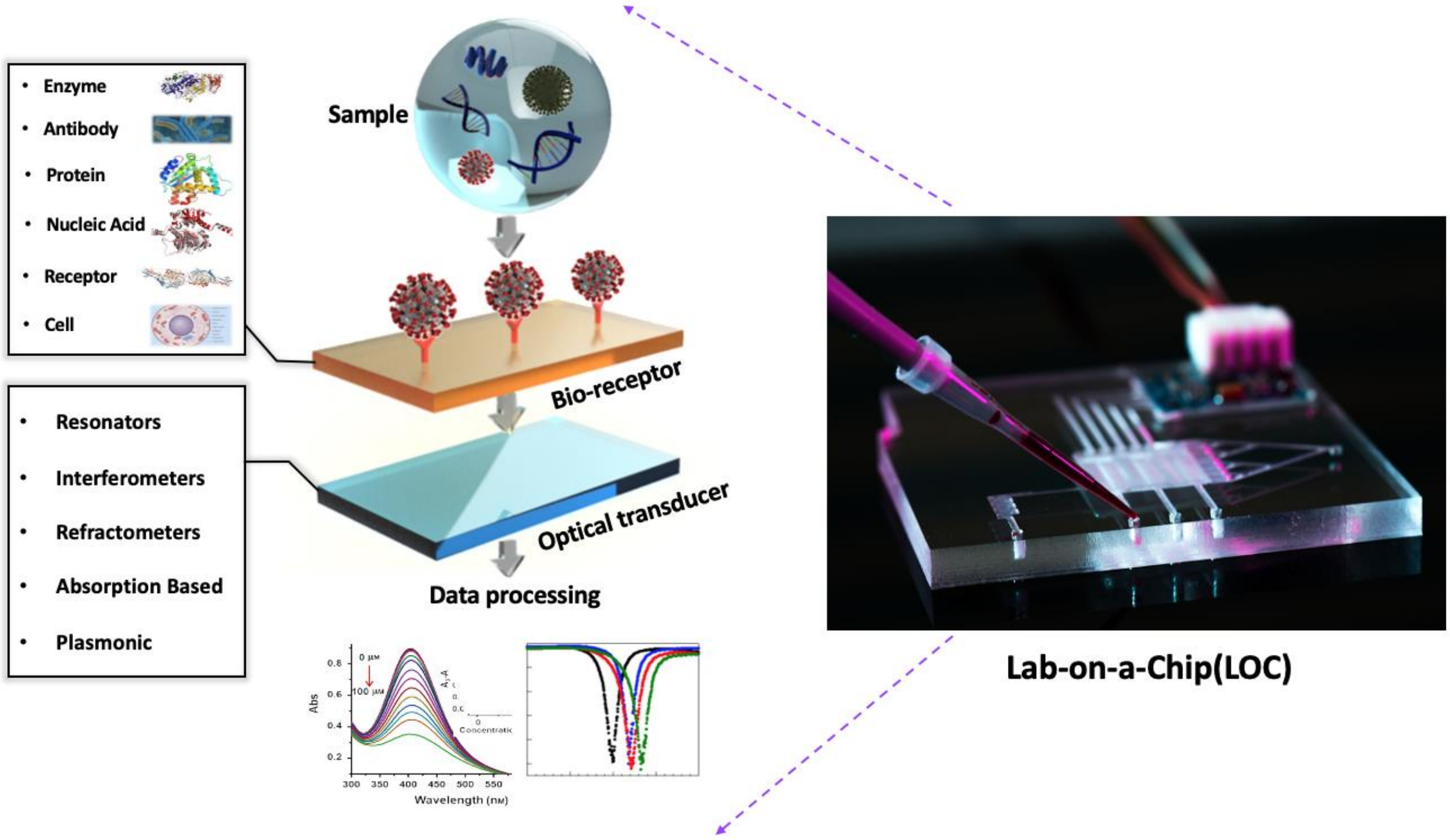

**Fig 2.** Schematic working principle of a lab-on-a-chip (LOC) optical biosensor.

Optical transducers have been extensively researched, commercialized and deployed in hospitals. Label-free optical biosensing can provide sensitive and durable point-of-care testing (POCT) device which is imperative for SARS-CoV-2 epidemic containment as it also can be easily operated at the point-of-risk by individuals without specialty training[40] (Fig. 2). On top of that, label-free optical biosensors have exhibited a strong conceivable potential to grow expeditiously in healthcare and biomedical fields as they provide a condensed accurate analytical tool to promote mass-scale screening of a broad range of samples through different parameters[35,41,42]. Optical biosensing does work in different physical transduction principles such as interferometers, resonators and plasmonic[43] and has been investigated to monitor many viruses with a good accuracy in different studies[31,44,45].

E. Surface functionalization strategies

The difficulty of measuring physical features of biological analytes in a biosensor has led to label-based techniques in which an additional molecule is attached to immobilized target molecules, viruses, or cells to enhance quantitative signal[41]. In a typical biosensor, the specific bioreceptors are immobilized on the chip sensing area to detect the targeted pathogens or proteins. Only the target biomolecules will be bounded to their corresponding biomolecular receptor upon introduction of analytes into the sensing area.

Notable examples of labels used in biosensing are dye molecules, a fluorescent tag, or enzyme. Various types of bioreceptors-targets coupling mechanisms have been also demonstrated in Fig 3 including antibody-antigen binding, enzyme-substrate catalytic reaction and cDNA-DNA hybridization. These labels require sophisticated reagent selection and modification that in turn come with the drawback of perturbing the assay and making final detection a challenging task. On top of that, labeling chemistry is both expensive and time-consuming. Thus, recent development in biosensing systems have been more intrigued by unlabeled or unmodified biomolecules (label-free biosensing)[35,41,46,47]in which native molecular properties like molecular weight and RI are utilized for sensing. Label-free detection does have its own shortcomings for example, it requires low non-specific binding[43] and sufficient signal to be generated upon targets binding[35]. However, its benefits such as providing real time analysis by simplifying assays and reducing time and number of steps required as well as eliminating experimental uncertainty far exceed its limitations[48].

The sensing transduction signals in Optical label-free biosensing platform functions based on miniscule changes in refractive index resulting from the attachment of biomolecules to the immobilized bioreceptors. It is of vital importance to have a highly sensitive biorecognition layer on the transducer surface in a label-free optical biosensors[43,49,50]. It goes without saying that the biosensor final sensitivity and specificity is strongly dependent on the immobilized molecules and the accessibility of target analytes to them. Therefore, the optimization of sensing surfaces and their biofunctionalization strategies is a significant factor for an accurate label-free optical biosensor where the sensitivity and accuracy are highly necessitated[43].

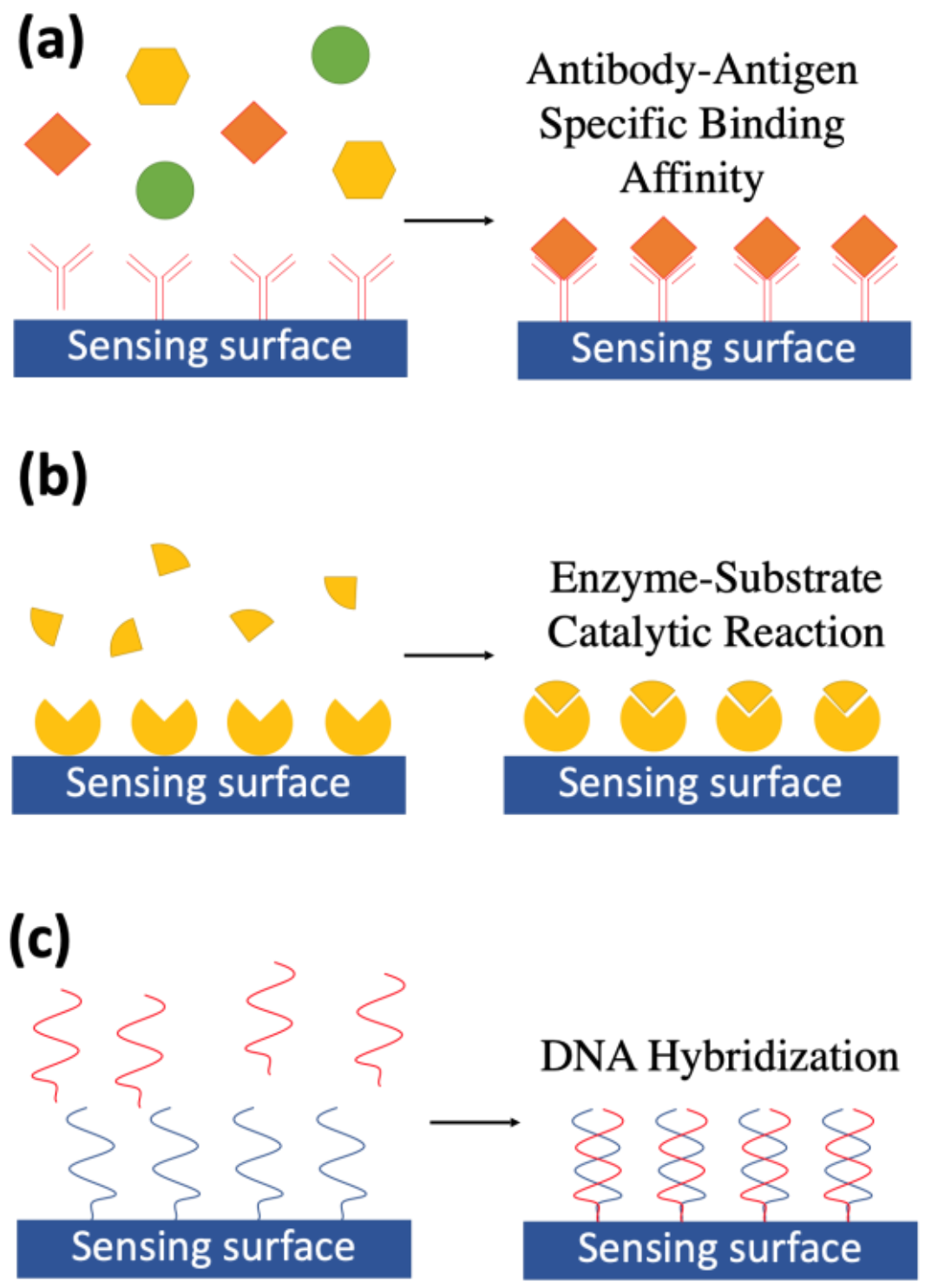

**Fig. 3.** Various receptors-targets coupling mechanisms on the biosensor surface showing (a): Antibody-Antigen binding b) Enzyme-surface catalytic reaction c) DNA hybridization d) schematic of SARS-CoV-2 virus binding to ACE2 receptors on a human cell (Inset shows the coronavirus spike protein (red) mediates the virus entry into host cells. It binds to the angiotensin converting enzyme 2 (blue) and fuses viral and host membranes).

The diverse range of target molecules and biosensor applications make it extremely difficult for a universal surface biofunctionalization procedure to be obtainable; hence, the procedure needs to be custom designed consequently. In a graphene-based Field Effect Biosensing(FEB), the surface is functionalized for protein immobilization with anti-Zika NS1 mouse mAb 6B1 developed by the Centers for Disease Control and Prevention[51]. Polyethylene glycol (PEG) has been studied particularly for removing proteins from the surface of silicon based biosensors as it provides a stable and anti-absorptive block against undesired non-specific interactions[52,53]. Detection of specific antibodies directed to viruses such as SARS-CoV-2 is another choice of identification for spotting infection. It is widely accepted that immunoglobulin M (IgM)and G (IgG) provide the first line of defense during viral infections[27]. However, it has been shown that for SARS infections, the IgM and IgG antibody could be detected in patients' blood after 3–6 and 8 days, respectively. Thus, targeting antibodies is not a suitable method for screening early cases of COVID-19. Detection of COVID-19 viruses in the range of picomolar (pM) to attomolar (aM) concentrations are possible if the biosensor provides a sensitivity range of $10^{-6} \sim 10^{-7}$ refractive index units (RIUs). Slow light-based biosensors are tremendously sensitive to refractive index variations and can determine the interactions with receptor biomolecules with the pM level. Most commercially available POCTs like home-use pregnancy and influenza test are based on immunostrips with the sensitivity level of nanomolar[54]. Both observation of the resonant shift or the reflectivity change will fulfill the requirement of the sensitivity.

The two-dimensional AuNIs functionalized with cDNA receptors (RdRp-COVID-C) can perform a selective detection of the RdRp-COVID through DNA hybridization[29]. The surface functionalization of sensor surfaces has resulted in sensitivity improvement and suppressing the nonspecific bindings[29]. The virus-like particle absorption has resulted in more SPR peak shift on non-functionalized surfaces compared to functionalized one simply as there are a greater number of adsorption sites on it[53].

Different structures have their own advantages and disadvantages. Elisa-based sensing requires additional secondary binding anti-body usage while it limits the application for fast detection requirements.

## II. INTEGRATED OPTICAL BIOSENSORS

Here, we explore the most well-defined bio-photonic sensing mechanisms based on functionalized waveguides, interferometer, and resonance shift in microcavities.

### A. Evanescent wave sensing: refractive index and absorption variation

Due to its immunity to electromagnetic interference (EMI), compactness and high selectivity, optical waveguides have attracted special attention as a basic biosensing system[55,56]. The interaction of target molecules with bioreceptors on the surface leads to effective index and absorption coefficient change (Fig. 4 (a)). The effective index change or loss change is a function of the concentration of biological or chemical targets on the surface. Effective mode index change for a perturbed waveguide can be calculated through the variation method[47]:

$$\Delta n_{eff} = \frac{n_m^2 - n_c^2}{Z_0 P} \iint |E(x,y)|^2 dxdy \tag{1}$$

where $E(x, y)$ is the electric field, $Z_0$ is the free space impedance and $P$ represents the light wave power, $n_c$ and $n_m$ are the refractive indexes of the aqueous solution without analyte, and molecular adsorption layer, respectively.

The vertical distance from the interface z does reduce the strength of the evanescent electric field exponentially. The penetration depth (d) can be calculated from the incident wavelength (λ) and incident angle ($\theta$) using the following formula:

$$d = \frac{\lambda}{4\pi \sqrt{n_0 sin^2\theta - n_1^2}} \quad (2)$$

Sensitivity (S) and limit of detection (LOD) are the two main criteria for evaluating the performance of the evanescent field sensor, which in turn depends on the strength of the interaction between the substance and light in a solution or on the surface[56]. In waveguide-based sensors, the expression of surface sensitivity ($S_{WG}$) is defined by the change in the effective refractive index ($\partial n_{eff}$) of the surface vis-a-vis the change in the additional layer of thin molecules ($\partial \rho$):

$$S_{WG} = \frac{\partial n_{eff}}{\partial \rho} \quad (3)$$

The surface sensitivity to specific target molecules is not suitable for general comparisons between sensors operating with different biosensing assays.

Figure 4 illustrates several different types of typical waveguide-based biosensors. Surface perturbation will cause change in absorption coefficient, which can be monitored in the output intensity (I) of waveguide. Optical waveguides have been extensively explored in different platforms including, but not limited to, rectangular waveguides[57], tapered fibres[58] and more sophisticated structures such as photonic crystal waveguides (PCW) [47,59–68], subwavelength grating[57,69] and Bragg grating[70]. As shown in Fig. 4 (b), Klimant et al. integrate a sensor-based planar waveguide with a 460 nm light emitting diode (LED), compatible with inkjet or screen-printing processing technologies. Correspondingly, the photodiode can be processed from the solution through printing technology, so that the sensor itself can be easily integrated into a single chip. The sensitivity of the optical sensor can be enhanced by extending the optical path. On the other hand, by simply depositing a gold layer, the device concept can be used for a surface plasmon resonance, which can improve the LOD even further[71]. Lear et al. reported an organic nanofilm based on optical waveguide biosensors (Fig. 4 (c)). The evanescent field is coupled to an integrated detector array that is buried beneath the waveguide. Toward the addition of nanofilm on the surface of the waveguide, the optical sensor exhibits a sensitivity of 20% modulation per nm. The device is capable of sensing multiple analytes simultaneously[72]. Another interesting approach have been introduced by Shi et al. They reported a variety of analytes with integrated fluorescence based on multi-channel sensors (see Fig. 4d). A broad linear response range was measured; 0.36 µg/L to 2.50 µg/L with a LOD of 0.21 µg/L is reported. This idea provides a method for measuring MC-LR in actual water samples[73].

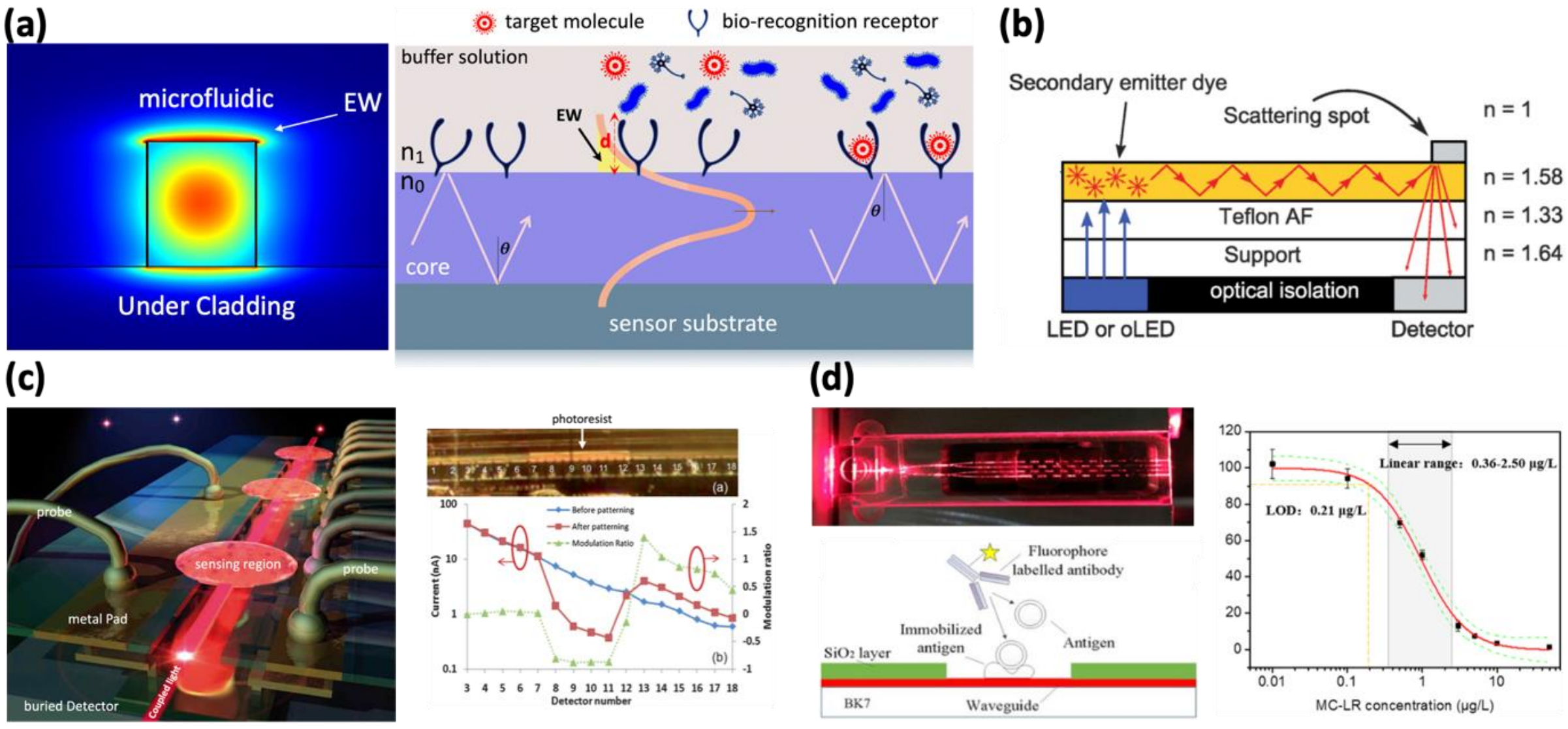

**Fig 4. (a)** Schematic of a waveguide-based biosensor facilitated through biorecognition element attached on its surface to bond with COVID-19 molecule. (b) Side view scheme of an optical sensor. This device provides an easy method to couple light into the planar waveguide free from prisms or gratings. The entire sensor can be fabricated by inkjet or screen-printing techniques[71]. (c) (left) Schematic view of the array coupled biosensing device working based on the field localization redistribution which happens when a waveguide cross-section changes during the specific binding of analyte in the targeted regions (right) Photocurrents measured before (blue line) and after (red line) the photoresist is patterned as well as the resulting modulation ratio (green line)[72]. (d) (top-left) Photographic image of light propagation along the waveguide-based sensor and (bottom-left) A cross-sectional view of the waveguide-based sensing, isolation layer and location of the surface chemistry (right) Using anti-MC-LR monoclonal antibody (MC-LR-MAb, 8C10), a linear dynamic response range of 0.36 µg/L to 2.50 µg/L with a LOD of 0.21 µg/L is determined, crucial for measurement of MC-LR in real water samples[73].

In order to maximize the sensitivity of the waveguide-based biosensor, the speed of light can be reduced even further. For this reason, the group velocity of the input pulse can be reduced by designing the photonic crystal structure[35]. Photonic crystal waveguides slow down the speed of light and introduce other enhancement factors, thereby enhancing absorption-based sensing on the surface of the waveguide.

B. Integrated interferometer sensing

Integrated interferometer photonic is one of the most practical architectures for sensing applications. It is based on splitting the input beam into two arms through a Y-junction, one arm is completely retained as the reference arm, and the other arm contains the target. The interaction of electromagnetic waves on the sensing arm will cause a phase difference with respect to the reference arm, and recombination of the two beams in the output will cause constructive or destructive interference, see Fig 5 (a).

Output intensity $I_{Out}$ of the MZI is described as follows [34,74,75]:

$$I_{Out} = I_{sen} + I_{ref} + 2\sqrt{I_{sen}I_{ref}}\cos(\varphi_0 + \Delta\varphi) \tag{4}$$

where $I_{ref}$ and $I_{sen}$ are light intensity in reference and sensing arms, $\varphi_0$ is initial phase difference between two arms without external perturbation. The sensitivity of the MZI-based sensor is related to the phase sensitivity relative to the length of the sensor arm:

$$S_{ph} = \frac{\Delta\varphi}{\Delta n_{eff} L} \tag{5}$$

In an imbalanced MZI, considering the phase matching condition, vis-a-vis the wavelength sensitivity, we can approximate the phase sensitivity of the MZI-based sensor:

$$S_{ph} = \frac{2\pi}{\Delta\lambda}\frac{S_{FSR}}{L} \tag{6}$$

where, $\Delta\lambda$ is the free spectral range (FSR) and $S_{FSR}$ is the spectral sensitivity.

Chemical and biosensing via Mach-Zehnder interferometers (MZI) have been widely exploited[76] and judicious design of imbalanced interferometers based on waveguides and photonic crystal waveguides explored by a lot of research groups[35,59,61]. Fig. 5b shows an MZI biosensor with a silicon nitride strip waveguide as the reference arm, and a silicon nitride slot waveguide as the sensing arm. Park et.al[76] designed a slot waveguide in the sensing arm to maximize the overlap between light and target analyte (see Fig. 5c). They reported a wide range of linear tests with concentrations ranging[77] from 19 fM to 190 nM, R2 = 0.979177. The asymmetry of the MZI array is used to detect miRNA in human urine samples, and its linear fitting curve is R2 = 0.997. Pavesi et. al[78] demonstrated detection of Aflatoxin M1 functionalized with antibody fragments. They demonstrated a volumetric sensitivity of 104 rad/RIU, leading to a LOD below $5 \times 10^{-7}$RIU (see Fig. 5d).

As shown in Fig. 5e, Morthier et. al have shown a polymer based on a lateral bimodal interferometer. They used two transverse modes in the waveguide to create the interferometer. The overall sensitivity of the manufactured interferometer sensor with a sensing length of 5 mm is reported to be 316π rad/RIU, and the extinction ratio can reach 18 dB. This method is promising for future commercial development. However, reliability of polymer needs to be further studied. Figure 5g shows the schematic structure of a spatially resolved resonant waveguide grating (RWG) for single cell detection. The sensor consists of (1) glass substrate, (2) a grating part, and (3) a waveguide with high refractive index. Owing to total internal reflection after the light guides in the waveguide, the light propagates out of the waveguide. Note that the wavelength of the resonance is unique to the coupled light and is sensitive to the local reflection index, which is proportional to the density of the target analyte in the penetration depth of the biosensor. The mechanism of RWG device can be also considered in the resonance-based biosensors. Figure 5 h,i illustrate a bimodal waveguide (BMW) sensor. BMW sensor is a universal path interferometric device based on the principle of the evanescent field detection. When we use it in mixture with a bio-recognition element it can directly detect the analyte of interest. The bimodal waveguide supports two transverse modes at the center, while single mode waveguide forms the input and output at the two sides of the bimodal waveguide. Various reports have shown sensitivity up-to tens of part-per million (ppm) level (Fig. 5h) for gas sensing as well as bio sensing (Fig. 5 f, i). Figure 5i, shows Si3N4 surface of a BMW device used to enhance the performance of the biosensor. The propagation mode is caused by the sudden increase of the waveguide core, which changes from single-mode operation (150 nm) to double-peak operation (340 nm). Fundamental mode operation relies more on the propagating wave at the core than the excited mode. Thus, both modes are affected differently through their evanescent field leading to waveguide output interference because of biorecognition event happening on the sensor area surface. This change can provide the sensitivity of the system. Analogous platform with a laser-based bimodal waveguide interferometer is proposed to detect the COVID-19 via changes in the sensor's evanescent light field (Fig.5 f).

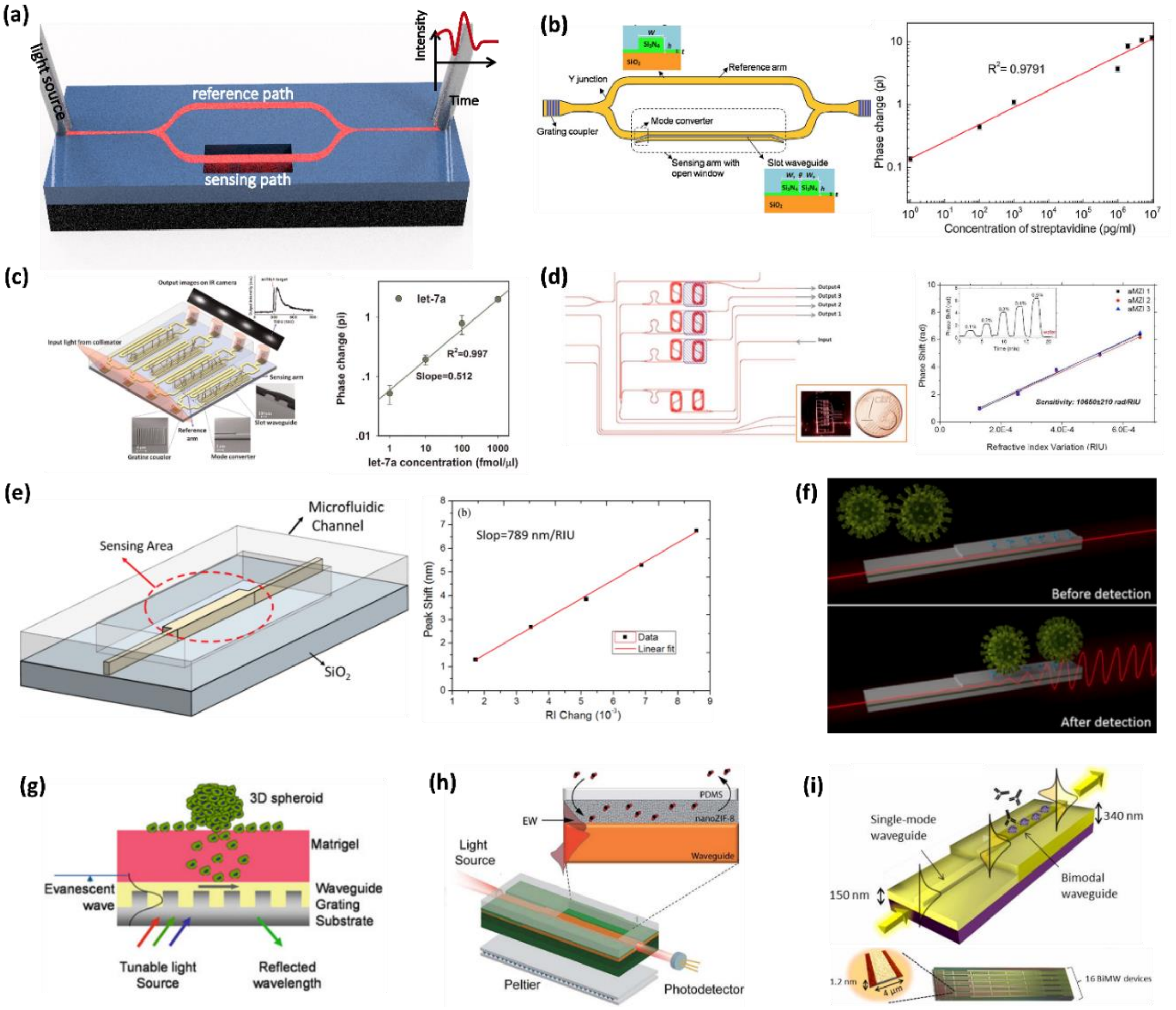

**Fig 5.** (a) Schematic diagram of MZI biosensor[76], The MZI biosensor includes a reference arm, and a sensing arm connected to two Y junctions. (b) (left) An MZI biosensor with a silicon nitride strip waveguide as the reference arm, and a silicon nitride slot waveguide as the sensing arm, (b) (right) shows the phase change induced by streptavidin binding as a function of the concentration with the range of 19 fM to 190 nM[76] which provides a linearity of R2= 0.9791, (c) (left) Schematic diagram of silicon nitride based on MZI[77], used to detect miRNA in human urine samples, (c) (right) Phase change to the binding of target miRNAs with a linear fitting curve of R2= 0.997, (d) (left) Schematics of an asymmetric MZI biosensor with purified solutions, (d) (right) Volume sensitivity measurements with ~ 10650 rad/RIU[78], (inset) phase shift curve for an MZI biosensor with different concentration of solutions. (e) Schematic structure of the polymer based bimodal interferometer. A 5 mm sensing length provides sensitivity of 316π rad/RIU, and extinction ratio of 18 dB. (f) A novel laser-based BMW sensor proposed to detect the COVID-19 via alteration in the evanescent light field. (g) Schematic structure of a spatially resolved resonant waveguide grating for single analyte detection. Due to the total internal reflection after the light guides in the waveguide, the light propagates and reflects out of the waveguide. Wavelength of the resonance is proportional to the coupled light and is highly sensitive to the local reflection index, which is directly related to the density of the target analyte within the penetration depth of the biosensor. (h) Schematic of the nanoZIF-8-based BMW sensor with the sensitivity level of tens of ppm, and (i) Interferometric mechanism of the BMW sensor. Further enhancement of the sensitivity is expected by introducing the slow-light effect into BMW sensors.

C. Resonance shift sensing

Contrasted to waveguide-based sensors that rely on light wave absorption, resonant displacement in functionalized microcavities provides a wide range of ultra-sensitive optical biosensors[79,80]. The magnitude of binding is determined by De Feijter's formula[81] that relates the absolute quantity of adsorbed molecules M with the change in the refractive index as:

$$M = d_A \frac{n_{A-}n_C}{dn/_{dt}} \quad (7)$$

where $d_A$ is the thickness of the adsorbed layer, $n_A$ is the refractive index of adsorbed molecules, $n_C$ is the refractive index of cover solution and dn/dc is the change in the refractive index of molecules, which is proportional to the shift dλ in position of the resonance peak. The size of the resonance wavelength shift is proportional to the number of adsorbed biomolecules, thus providing a label-free method to quantitatively determine the target analyte.

*1. Ring Resonators*

Although high quality (Q) ring resonators can be achieved with a larger radius, the trade-off between the Q and the free spectral range (FSR) limits the radius for a given FSR, which should be large enough for effective recognition of the sensing signal from the adjacent interference signals or for large scale on-chip multiplexing sensing applications. Wang et al. [82,83] proved through experiments that the Q-enhanced SWGMR was specially designed using a trapezoidal silicon column (T-SWGMR)( Fig 6). According to the report, the SWGMR has a Q of ~5600 even with a large radius of 15 µm, smaller radius provide much higher Q. Contrasted with conventional rectangular silicon pillars comprised of SWGMRs (R-SWGMRs), an asymmetric effective refractive index distribution is created, which can significantly reduce bending loss and thus increase the Q of SWGMRs.

The experimental results show that the applicable Q value of T-SWGMR with a radius of 5 µm is as high as 11,500, which is 4.6 times the Q value (about 2800) provided by R-SWGMR with the same radius, indicating that the propagation loss is reduced by 81.4%. To go one step further, Yan et al.[84] proposed a T-SWGMR biosensor and demonstrated the unique stable surface sensing characteristics through a demonstration of miRNA detection at a concentration of 1 nm (Fig 6b)

In addition to utilizing the unique stable sensing characteristics of SWGMR and the enhanced Q of T-SWGMR, Chang et al. [69] showed a pedestal T-SWGMR biosensor that maximizes the mode volume overlap by implementing an asymmetric refractive index distribution along the vertical direction on the silicon-on-insulator (SOI) platform, thereby further improving sensitivity( Fig 6.d). Both theoretic analysis and experimental proofs show that the volume sensitivity and surface sensitivity have been significantly increased by 28.8% and 1000 times, respectively. For streptavidin, a spectrometer with a resolution of 0.01 nm is used, and its LOD is about 400 fM. Owing to imperfect manufacturing process, experimental Q estimate of T-SWGMR with a radius of 10 μm and FSR of ~13 nm is 1800. The optimized SWGMR with symmetric coupling demonstrated by Huang; et al. [85] estimated Q to be 9800.

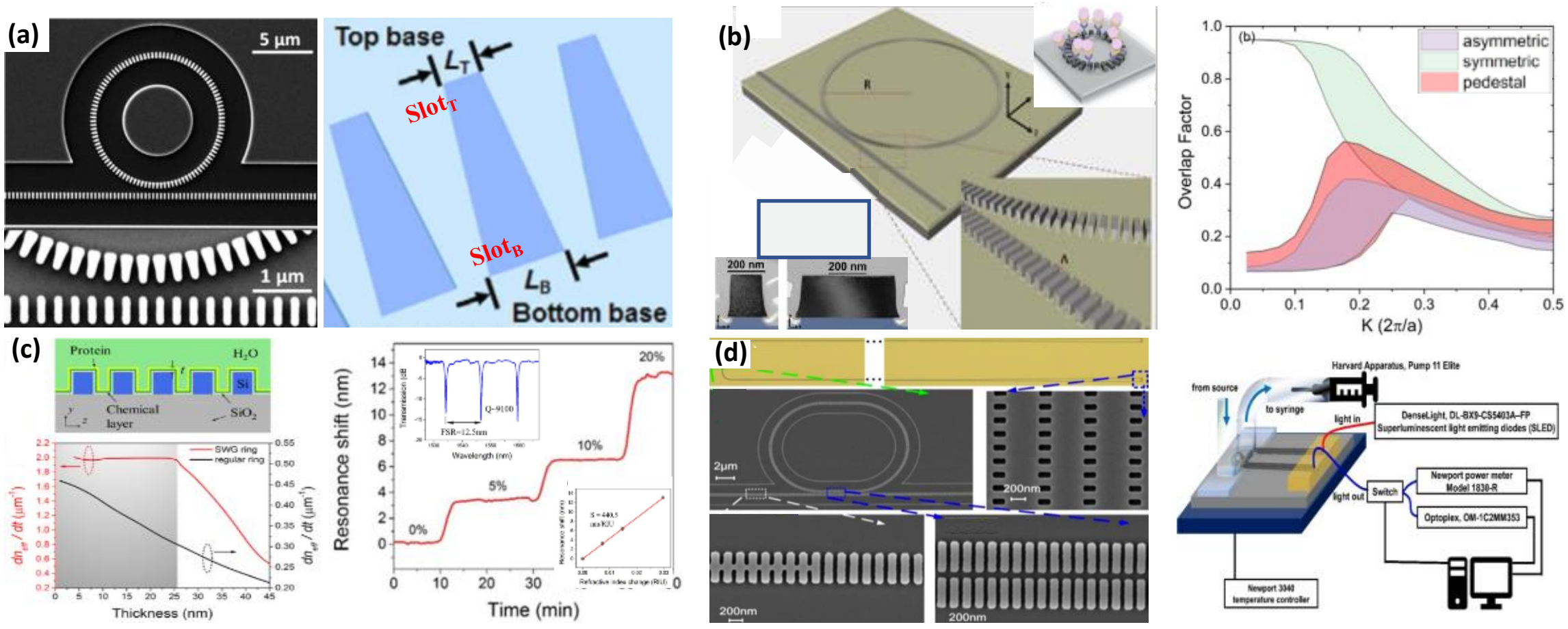

**Fig 6.** Subwavelength grating waveguide micro-ring/racetrack biosensors, (a) High Q subwavelength grating waveguide ring (SWGR) resonator based on trapezoidal pillars[82,83], (b) Pedestal SMGMR base biosensor for improved S[84], (c) T-SWGMR based high Q and high S biosensor[83], (d) Racetrack SMGR biosensor with high Q and high S[69].

## *2. Microtoroid*

Microtoroids are resonators with a Q of >$10^8$ and a small mode volume which can be fabricated on silicon using standard microelectronics techniques[86]. However, microtoroids need to be strictly aligned with the tapered fiber waveguide to achieve high coupling and cannot meet our needs for high-throughput multiplexing sensing. Vahala; et al.[86] demonstrates the possibility of detecting unlabeled single molecules and higher concentrations on a single platform (Fig 7a, b). According to reports, the quality of performing planar lithography is about $1.83\times10^8$. The author reports that by using the IL-2 solution, the micro-ring sensor can provide a dose response of $10^{-19}$M to $10^{-6}$ M and a working range of 5 aM to 1 μM. In another report for a microtoroid with a diameter of 90 μm, authors reported[87] that a measurement lifetime of 43 ns corresponds to an inherent quality factor of $1.25\times10^8$ (Fig 7c, d).

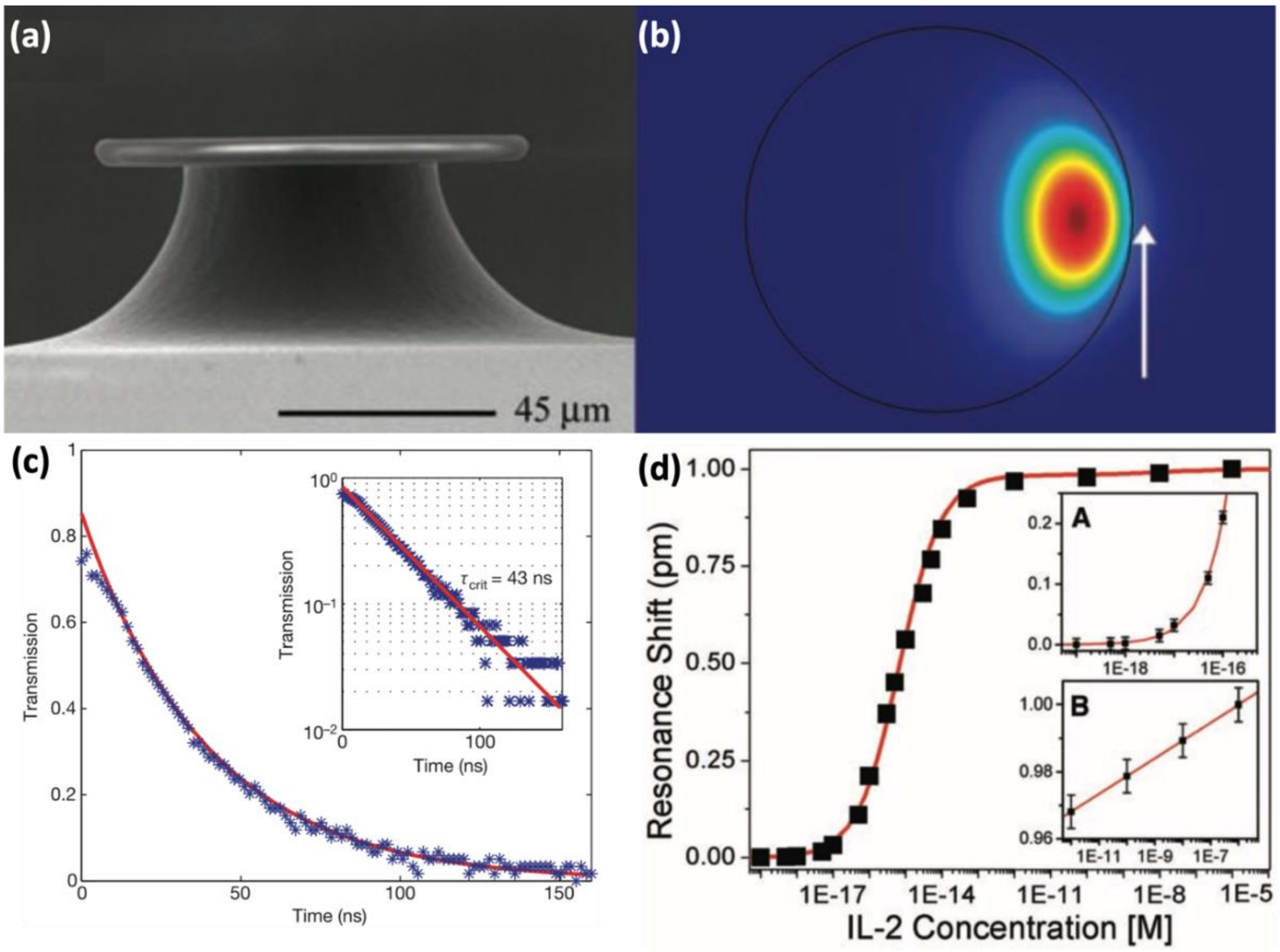

**Fig 7.** The cross-sectional view of the fabricated microtoroid based biosensor[86]. (a) SEM image of the UHQ microtoroid optical resonator, (b) A finite element model of a 4-mm minor diameter microtoroid resonator surrounded by water. Note part of the field leaks into the environment (white arrow). This interaction between the whisper gallery mode and the environment provides ultra-sensitive detection,(c) Measured lifetime of 43 ns[87] corresponds to an inherent Q of ~1.25 ×$10^8$, (d) Microtoroid with the quality of ~1.83 × $10^8$ and use of IL-2 solutions can provide a dose response of $10^{-19}$ M to $10^{-6}$ M and a working range of 5 aM to 1 µM.

### *3. Photonic Crystal*

For micro-cavity photonics crystal (PhC) sensors, we mainly use 2d-PhC biosensors, which have the advantages of design flexibility, compact size (surface area of about a few square microns) and strong light interaction with the analyte of interest. As shown in Fig. 8 a-b, Chen; et al. recently confirmed the work of practical pancreatic cancer detection by using nanopore-assisted high-Q (22000) and high-S (112nm/RIU) L13 PhC cavities[60]. The detection results show that a concentration of 8.8 femto-molar (0.334 pg/mL) pancreatic cancer biomarker was successfully detected in patient plasma samples, which is 50 times more diluted than conventional enzyme-linked immunosorbent assay (ELISA). To go further, by designing and developing multimode interference (MMI) separators[47,62,64–66,88], high-throughput and multiplexed biosensor arrays have been proposed and demonstrated, (Fig. 8 a, b[47,64–66]). The integrated scheme and array method proposed and proven improve the multi-parameter and multi-function detection capabilities of the sensor and can be used in practical diagnostic applications.

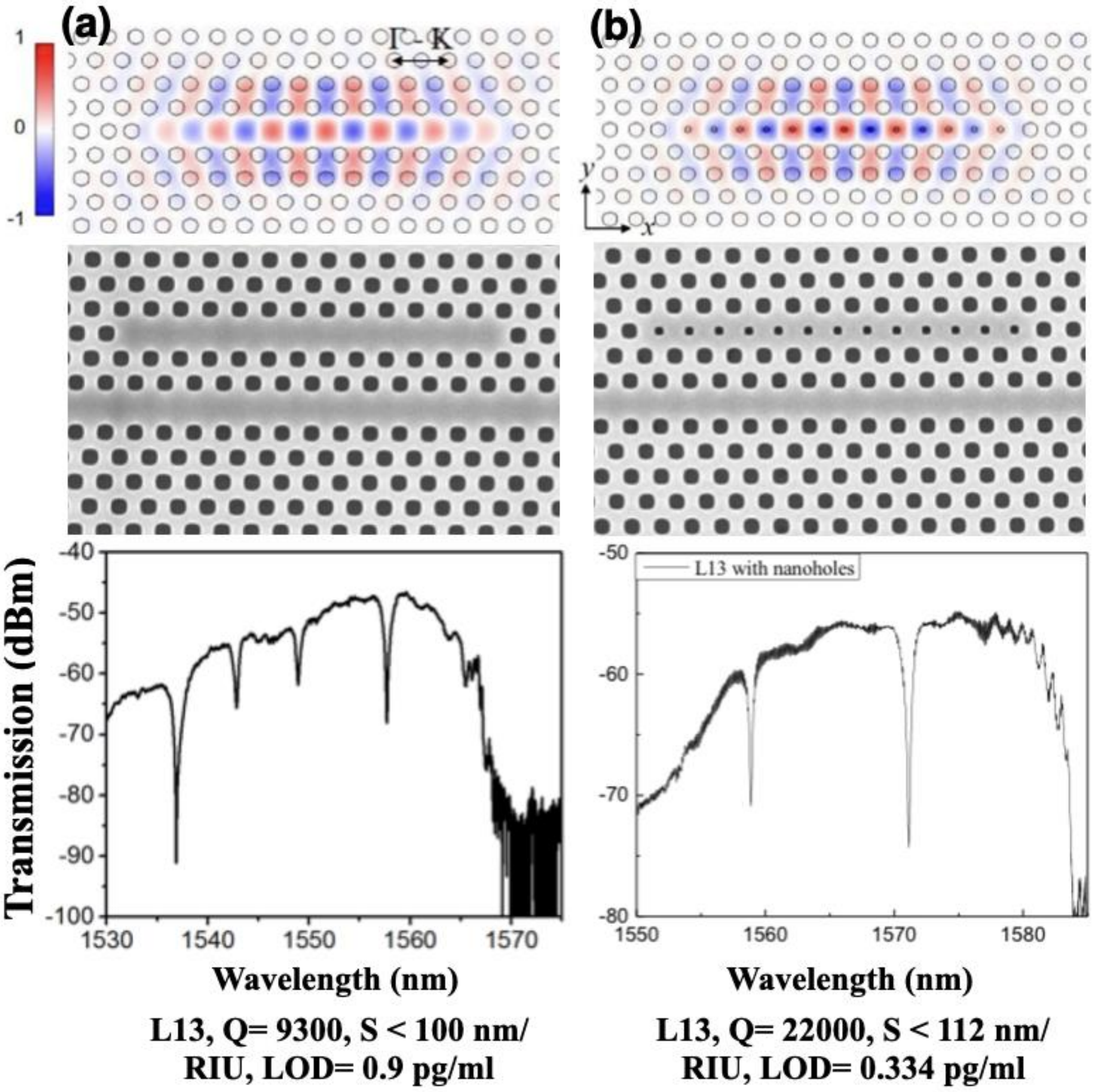

**Fig 8.** 2D-PhC micro-cavity biosensors. (a) Slow light L13 PhC cavity for enhanced sensitivity[68] ; (b) Nano-holes assisted high sensitivity L13 PhC cavity for plasma protein detection of pancreatic cancer[60]. The Q, S and LOD performances are annotated below each picture. Note that there is a trade-off between sensitivity and LOD.

Our team has extensively described the functionalization of silicon surfaces using various probe biomarkers in past research[47, 64-66]and their use in the detection of specific conjugated biomarkers using our silicon photonic crystal microarray structure. Previously, we demonstrated a multimode interference coupler architecture shown in Fig. 9a,b also shows the series and parallel integration of 64 sensors on the silicon chip. Multiplexed sensing with specificity of lung cancer cell line lysates were demonstrated. We have also demonstrated experimentally that the silicon photonic crystal sensor chips can be fabricated in a commercial foundry for high volume manufacturing. For COVID-19 testing, the silicon chip manufacturing process and sensor functionalization process will be the same as before.

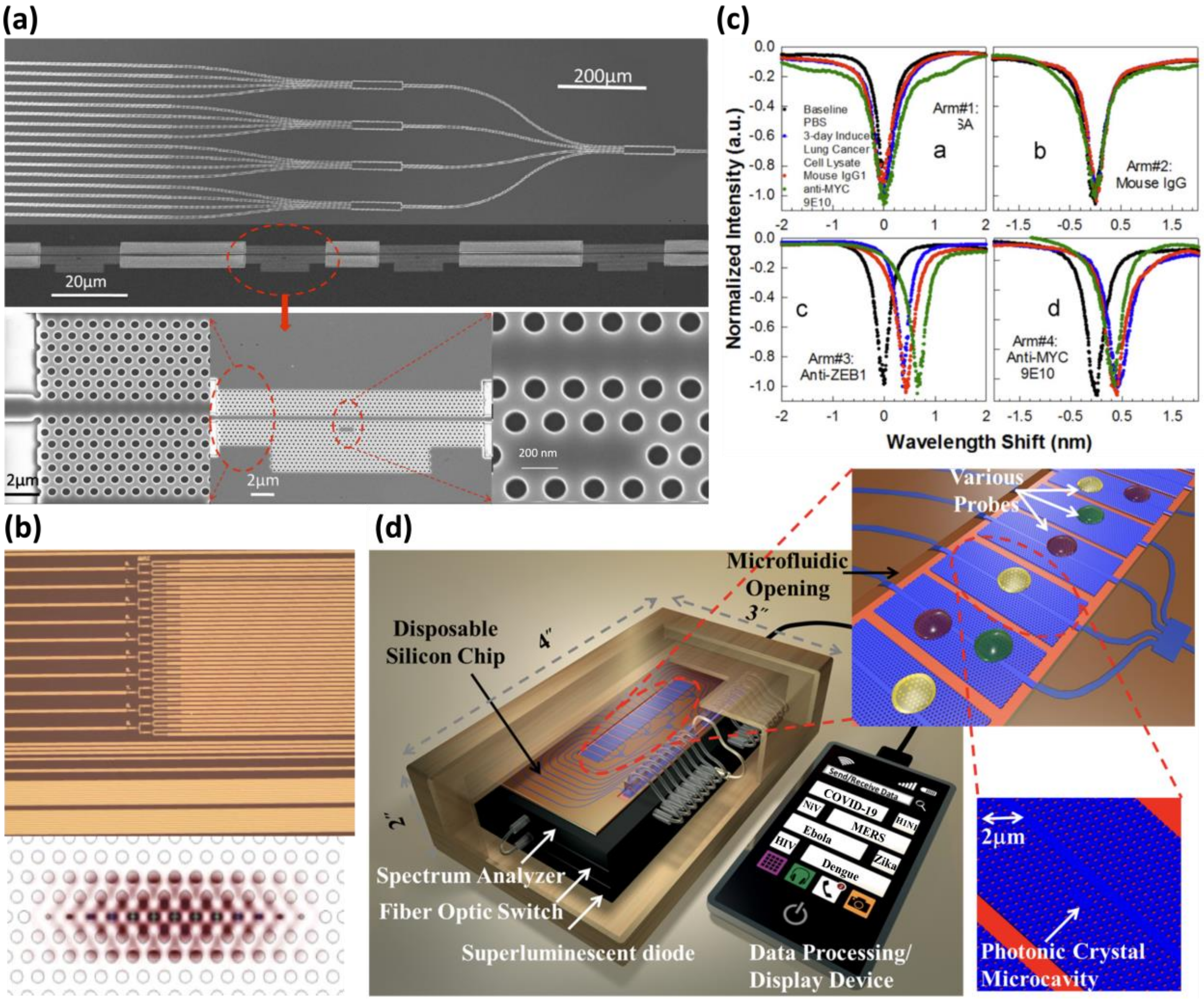

**Fig 9.** Integrated sensor array for high throughput detection[47, 64-66] (a) Multiplexed 1×4 MMI power splitter that splits an input light into 16 optical paths, each with 4 photonic crystal microcavity sensors for 64 sensors in total, (b) (top) Microscope image of foundry fabricated silicon photonic crystal sensor devices, (bottom) Highly confined electric field in a photonic crystal microcavity for enhanced analyte sensitivity.(c) Multiplexed simultaneous specific detection of ZEB1 in lung cancer cell lysates with four arms of the MMI derivatized with bovine serum albumin , isotype matched control mouse IgG1, anti-ZEB1 antibody and anti-MYC 9E10 antibody, and (d) Comprehensive PCW detection platform.

Label-free microarrays are particularly exciting because it simplifies biochemistry significantly when probe-target binding conjugations can be studied without steric hindrance associated with fluorescent or radioactive tags. In Fig. 10, we compare our photonic crystal microarray approach with other research performed using photonic crystals (PCs) and show that our microarray has the highest sensitivity to small changes in concentration. It summarizes sensitivities and detection limits demonstrated in our system compared to other label-free methods, including surface plasmon resonance (SPR), opto-fluidic ring resonators (OFRR), ring resonator (RR) Pedestal SMGMR and photonic crystal (PC) devices, as function of sensing area. Sensitivities of PC microcavity structures demonstrated at Omega Optics (OO) and University of Texas (UT), Austin is indicated by UT/OO.

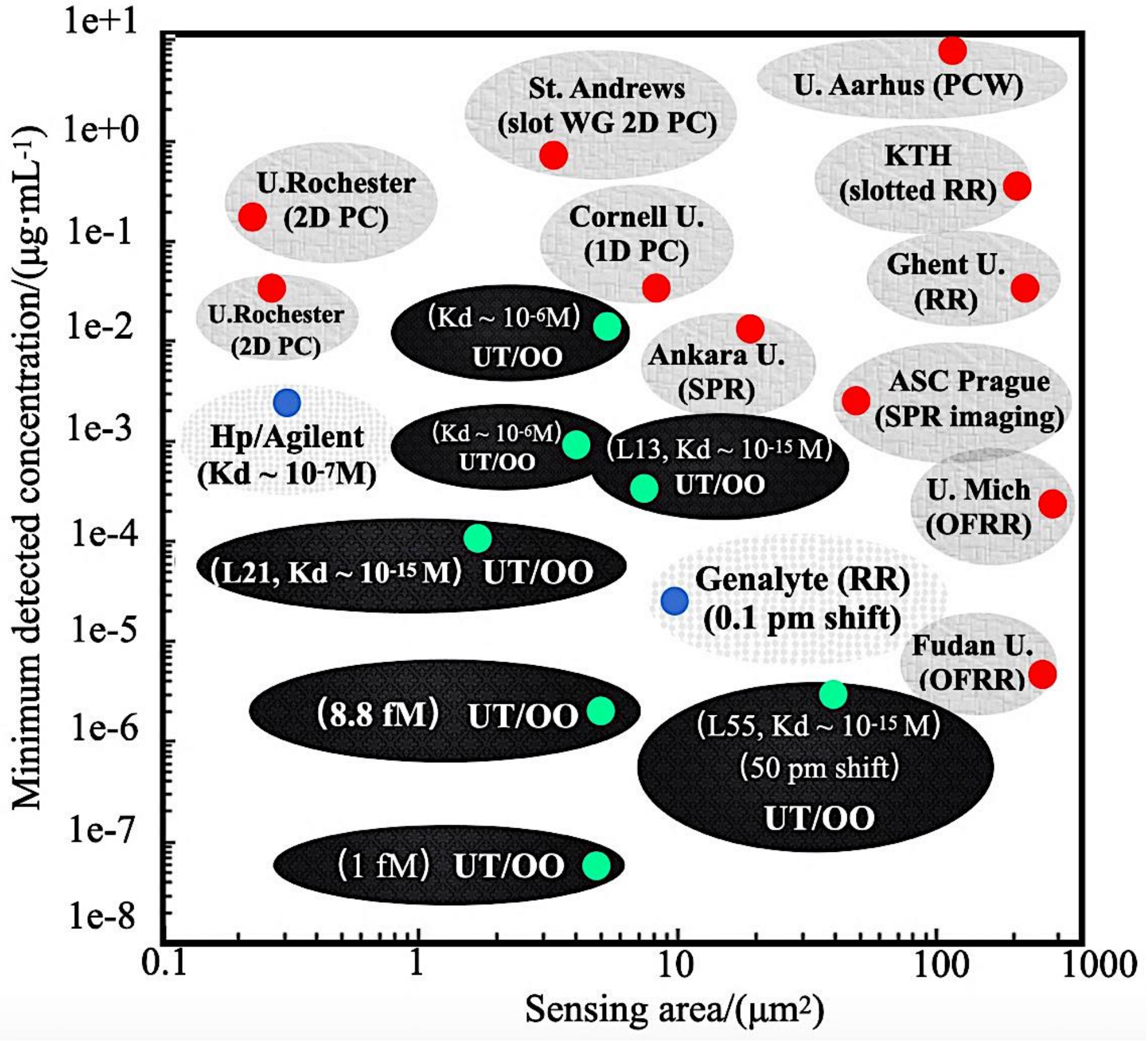

**Fig 10.** Comparison of minimum experimentally demonstrated detection limits versus other label-free optical platforms as a function of sensing area on chip. (Legends: PC=photonic crystal[47,59,61–66,68,69,84,85,88–92]; RR=ring resonator[69,93–95]; SPR=surface plasmon resonance[31,96] ; OFRR=opto-fluidic ring resonator[97,98], LCR=liquid crystal sensors[99], BIND=Bio-molecular interaction detection.) UT/OO[a] has detected 67 pg/ml which is the highest sensitivity reported. [a] UT/OO denoted in the figure is in reference to published works jointly by UT Austin and Omega Optics, Inc.

Our photonic crystal microcavity not only has high sensitivity and low detection limit, but also can achieve dense integration of sensors due to its small geometric size. In Table 1 below, we compare our proposed PC microarray platform with commercially available bench-top systems for water monitoring such as inductively coupled plasma mass spectrometry (ICP-MS) and inductively coupled plasma optical emission spectrometry (ICP-OES, for metals), gas chromatography–mass spectrometry (GC-and GC-MS for organics) and ELISA (for all analytes with bio-signatures). We compared the technical advantages of our proposed platform with other platforms and showed that our platform can provide comparable sensitivity to existing desktop systems, while also being portable.

**Table 1.** Summary of technical advantages of our work vs commercialized approaches [90–93,95–97,99–101 59,60,63,68,69,102].

| Technique | ELISA (all analytes) | Flame & GF AAS(metals) | ICP-MS; ICP- OES (metals) | GC (organics) | GC-MS (organics) | Our work |
|---|---|---|---|---|---|---|
| Label-free | No | Yes | Yes | Yes | Yes | Yes |
| Probe/Pretreatment Required | 100 µg-1 mg/spot | Ashing/acid pretreatment to remove organics | Ashing/acid pretreatment to remove organics | Sample extraction and clean-up | Sample extraction and clean-up | < 18 ag/spot |
| Sensitivity | Sub-nM for competitive ELISA | 0.002-1.5 ppb (depends on metal) | 0.2-0.0007 ppb (depends on metal) | Low ppb, depending on sample | Ppb to ppt, depending on sample | 67 fg/ml, for protein |
| Binding spots or samples analyzed | 96-384 Wells/ plate | 1 sample, sequential runs, one analyte | 1 sample, sequential runs, multiple analytes | 1 sample, sequential runs, multiple analytes | 1 sample, sequential runs, multiple analytes | Up to 128 Analytes (on one chip) |
| Platform | Affinity tags | Furnace, monochromator detector | Metals -> ions, then detected by MS or OES | Column separation in vapor phase | Like GC, then MS and detection | 2D PC |

## III. PLASMONIC OPTICAL BIOSENSORS

A plasmon can be described as a collective oscillation of a free electron or a quantum of plasma oscillation. Propagation of electromagnetic waves along the surface of a metallic surface or surface plasmons (SP) can be understood as a strong interaction between conduction electrons of the metallic surface and electromagnetic waves, which leads to resonance modes trapped on the surface, also known as surface plasmon resonances (SPRs)[31,44,45]. SPR propagation along the conductor surface produces a charge density distribution, which enhances the light matter interaction on the nanoscale. Such enhancements, so called “hot spots”, occurred at the interface between a dielectric and metallic surface offer the higher sensitivity for plasmonic biosensors. Many types of optical biosensors based on plasmonic platforms have been studied as fascinating candidates for biomedical and chemical sensors [28,29,46,103–113]. The selectivity of the plasmonic-based biosensor can be achieved by using immobilization of the various bioreceptors; depending on the target analytes, specific bioreceptors can be selected to be immobilized on the surface of the sensor and react or bind only to its counterparts. Based on the device configurations, plasmonic biosensors can be divided into two groups, SPRs and localized SPRs (LSPRs).

### *A. Surface plasmonic resonance (SPR) sensor*

Fundamentally, when the phase matching condition between the incident light and the SP wave guided along the metal/dielectric interface is reached, the incident light can be coupled to the surface guided mode. Note that the resonance condition between the incident light and the conductive electrons at the metal/dielectric interface with a fixed angle of incidence is only achieved at a specific wavelength. The guided light will be absorbed by the conducting electrons that resonates, which will significantly reduce the reflected light at that particular wavelength. Therefore, once the target molecule is attached to the functionalized metal film, the refractive index does change, causing a shift in the resonance wavelength. Consequently, SPR angle alteration can be characterized as the main sensing mechanism. Several coupling methods have been proposed, including a grating coupler, a waveguide coupler, and a prism coupler, but the prism coupling method has been used as a standard configuration based on the Kreichman configuration[114]. Figure 11. shows the schematic of the conventional SPR sensor configuration.

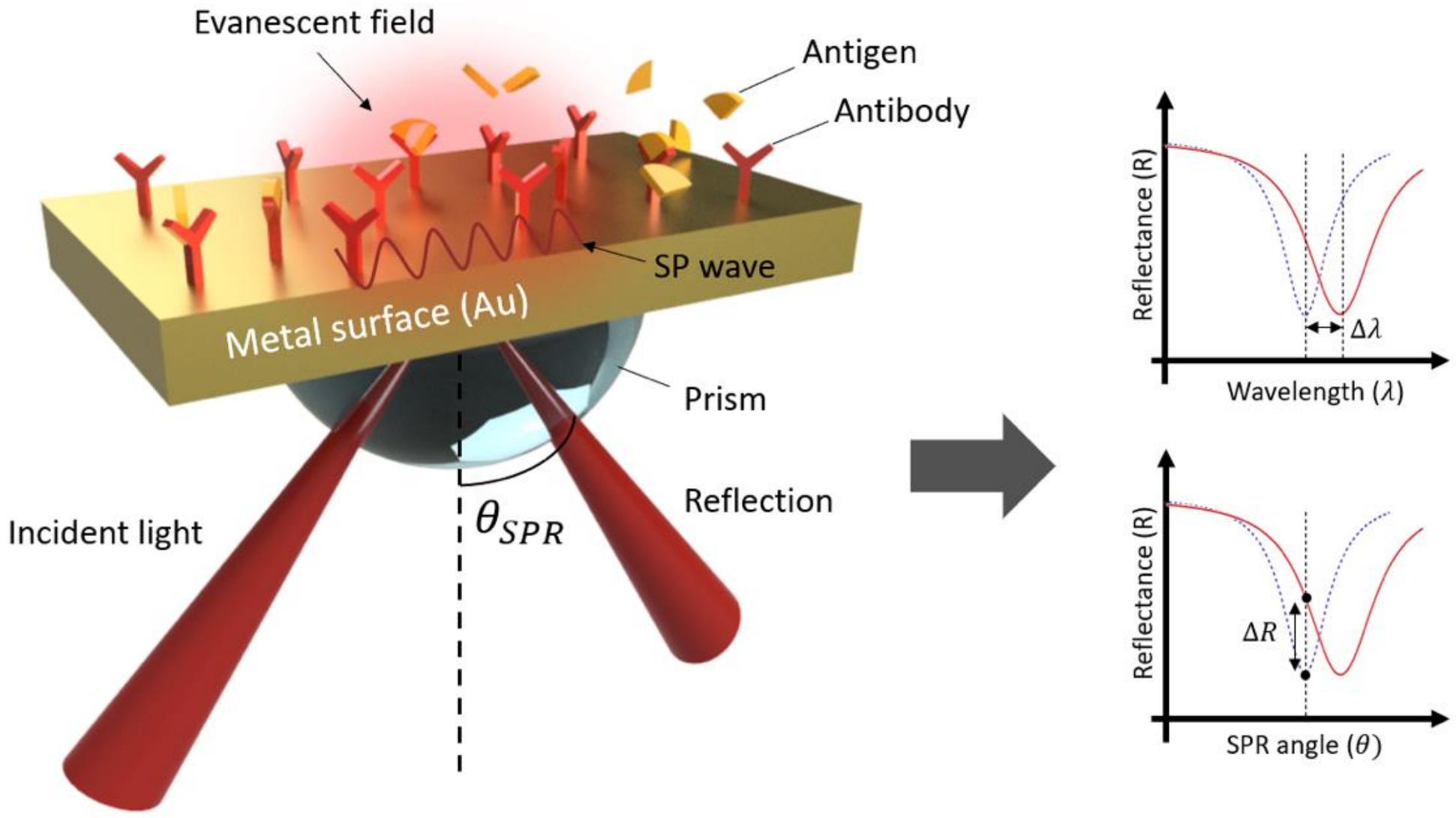

**Fig 11**. The schematic illustrations of the standard SPR based biosensor configuration.

The wavevector of the evanescent field of the incident electromagnetic wave propagating along the prism-metal interface is shown in the following equation[115,116] :

$$k_{in} = \frac{2\pi}{\lambda} n_p \sin(\theta) \quad (8)$$

where $n_p$ is the refractive index of the prism, $\lambda$ is the wavelength of the incident light, $\theta$ is the incident angle. The wavevector of the SP wave propagating along the metal/dielectric interface is as below[116,117] :

$$k_{SP} = \frac{\omega}{c}\sqrt{\frac{{n_D}^2{n_M}^2}{{n_D}^2+{n_M}^2}} \quad (9)$$

where $\omega$ is the angular frequency of the wave, c is the speed of light in vacuum, $n_M$ and $n_D$ are the refractive indices of the metal and dielectric. As aforementioned, the resonance condition is met when $k_{in} = k_{SP}$, so we can calculate the SPR angle in the following equation:

$$\theta_{SPR} = sin^{-1}(\frac{1}{n_p}\sqrt{\frac{{n_D}^2{n_M}^2}{{n_D}^2+{n_M}^2}}) \quad (10)$$

The sensitivity of the SPR devices are determined by the resonance shift with respect to the change of the refractive in the absence and presence of the target analyte [118–122]:

$$S=\frac{\Delta\lambda}{\Delta n} \quad (11)$$

where $\Delta\lambda$ is the resonance wavelength shift and $\Delta n$ is the change of bulk refractive index including the target analyte.

*B. Localized SPR sensor*

On the other hand, nanostructures in conductive thin films are among the essential building blocks of LSPR plasmonic biosensors (see Fig.12). These nanoscale geometric/periodic lattice factors bring huge advantages over conventional SPR devices. Contrasted with SPR occurring along the propagation surface, the attenuation length of the local electromagnetic field is much shorter. These strict restrictions, with a shorter sub-wavelength structure, can achieve ultra-low mode volume resonance, making it sensitive to environmental refractive index changes, which are particularly helpful for the detection of tiny biological molecules. Also, an incident light can be directly coupled to SP wave on the conductive structures without any external couplers, e.g., prism or gratings, which ameliorates the complexity of the entire system and enables the sensor miniaturization[123] and the absorbance, transmittance, and reflectance-based sensing[42,107–109,124–126].

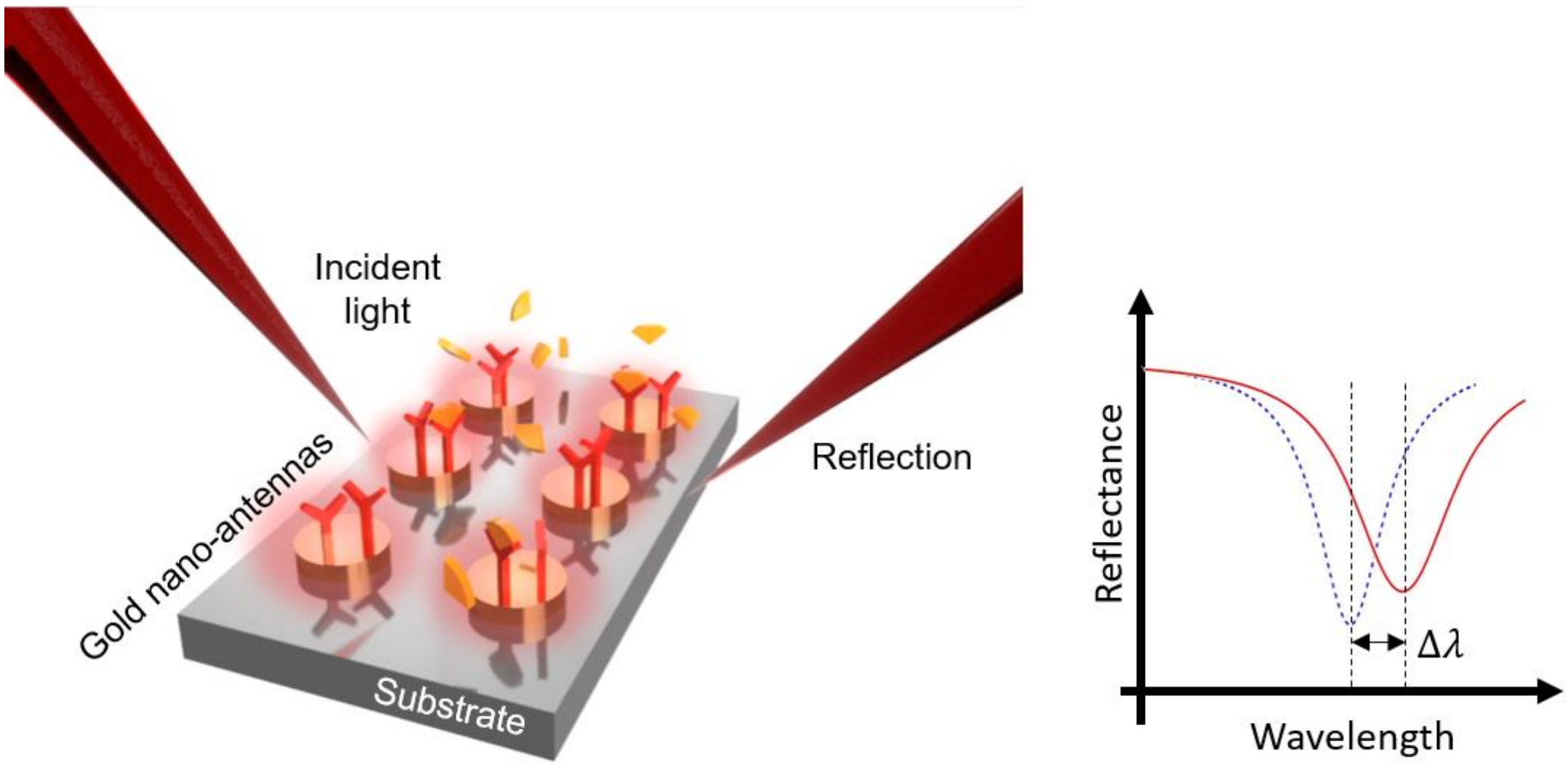

**Fig 12**. The schematic illustration of the resonance wavelength shift sensing based LSPR sensor configuration.

Moreover, LSPRs can be utilized for various types of resonance modes and detection methods, including surface-enhanced Raman spectroscopy (SER)[28,112], photo-luminescence/fluorescence[50,111], and mid-infrared spectroscopy[106], by tuning the resonance wavelength for a specific light-matter interaction. Table 2 shows the comprehensive comparison between the conventional SPR and LSRP biosensors.

**Table 2.** The comparison between SPR and LSPR sensors.

| | SPR | LSPR |
|---|---|---|
| Sensing distance | ~ 1000 nm | ~ 10 nm (tunable) |
| Coupling components | Required (prism, gratings and else) | Not required |
| Sensor miniaturization | Limited | Effective |
| Detection methods | Angle shift, Wavelength shift | Wavelength shift, extinction, scattering, imaging, SEIRA, SER, fluorescence, photoluminescence |
| Label-free detection | Yes | Yes |
| Respond time (Real-time detection) | $< 10^3$ s | $< 10^3$ s |
| Specificity | Achieved by surface functionalization | Achieved by surface functionalization, SEIRA offers the identification of molecules chemical bonds |
| Multiple microfluidic channels compatibility (Parallel detection) | Limited | Yes |

*1. Resonance shift sensing*

Figure 12 shows the basic configuration of the LSPR device based on the resonance shift sensing. The metal nanostructure on the dielectric substrate is used as a resonator, and due to the above-mentioned advantages, the sensitivity can be further improved compared with the conventional SPR resonance shift device. The sensitivity and the figure of merit of LSPR resonance shift sensors follows the same definition of SPR's as in eq. (11). Another important performance factor of resonance-based sensor is the quality factor (Q), which is defined as [127,128] :

$$Q = \frac{\lambda_o}{FWHM} \tag{12}$$

where $\lambda_o$ and FWHM are the wavelength and full-width half maximum of the resonance peak, respectively. To enhance the sensing performance, a higher Q value is desirable because of the reason that sharper peaks with high Q values are much easier to detect. Considering all these factors, the inherent detection limit (ILOD) of the resonance displacement sensing device can be defined as follows[80,129] :

$$\mathrm{ILOD} = \frac{\lambda_o}{Q \cdot \mathrm{S}} \tag{13}$$

which indicates that both the higher sensitivity (S) and Q factor are required to minimize the limit of detection of the sensors.

Although these so-called hot spots provide higher sensitivity for LSPR biosensors, their performance is greatly limited due to the basic limiting factors of ohmic losses in metal surfaces. In other words, compared to other photonic biosensors, the absorption loss in the conductive nanocavity leads to a low Q value, so research has been conducted to achieve low-loss devices by using advanced materials or optimizing the geometry of metamaterials.

*2. Plasmonic perfect absorber*

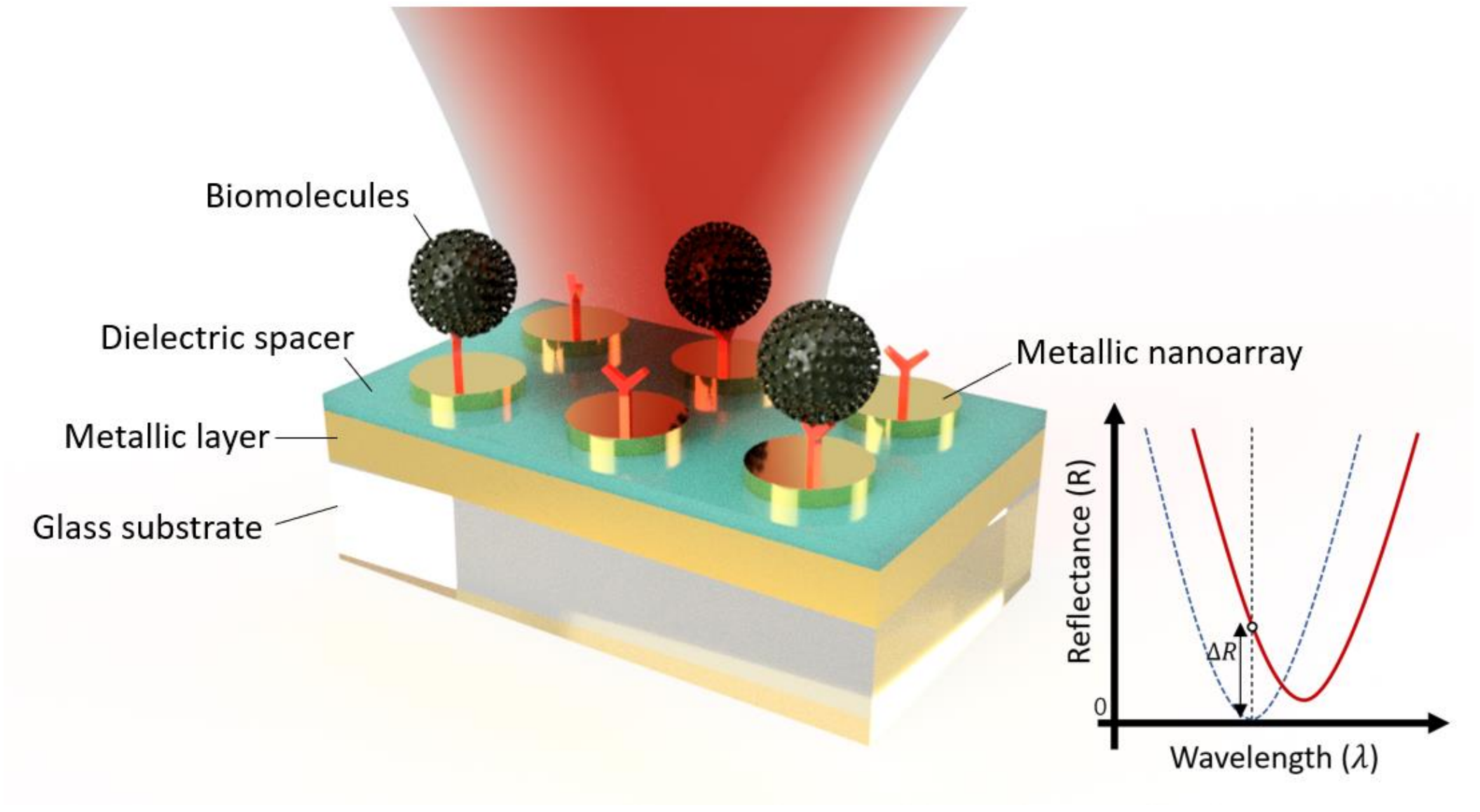

**Fig 13**. The schematic illustration of LSPR perfect absorber-based sensor.

On the other hand, the concept of a plasma perfect absorber (PPA) sensor was introduced to overcome this intrinsic limiting factor[130–134]. Figure 13 shows the typical configuration of PPA sensor consists of periodically arranged metallic nano antennas (metamaterial) on top and thin metallic 'mirror' layer on the bottom separated by dielectric spacer[132]. The basic concept is to have a perfect absorbance at the operating wavelength and make a 'zero' transmittance by maximizing the metamaterial losses; in other words, the losses are served as an advantage in the PPA sensors. In this structure, most of the incident light at the operating wavelength is absorbed by top nano antennas operating as a resonator through impedance matching, and the metallic bottom layer act as a 'mirror' to eliminate the transmittance. As a result, the reflectance of light can be characterized for sensing as in figure 13, and the figure of merit (FOM$_{PPA}$) is defined as below[133] :

$$FOM_{PPA} = \left| \frac{dI(\lambda_o)/I(\lambda_o)}{dn(\lambda_o)} \right| \quad (14)$$

where $dI(\lambda_o)/I(\lambda_o)$ is the relative intensity change of reflected light at a fixed resonance wavelength $\lambda_o$, which is induced by a refractive index change $dn(\lambda_o)$.

Moreover, it has been shown that the perfect absorption (>99%) of incident light at working wavelength can be remained over a wide incident angle and insensitive to the polarization (TE/TM) of incident light[133].

*3. Surface-enhanced infrared absorption spectroscopy*

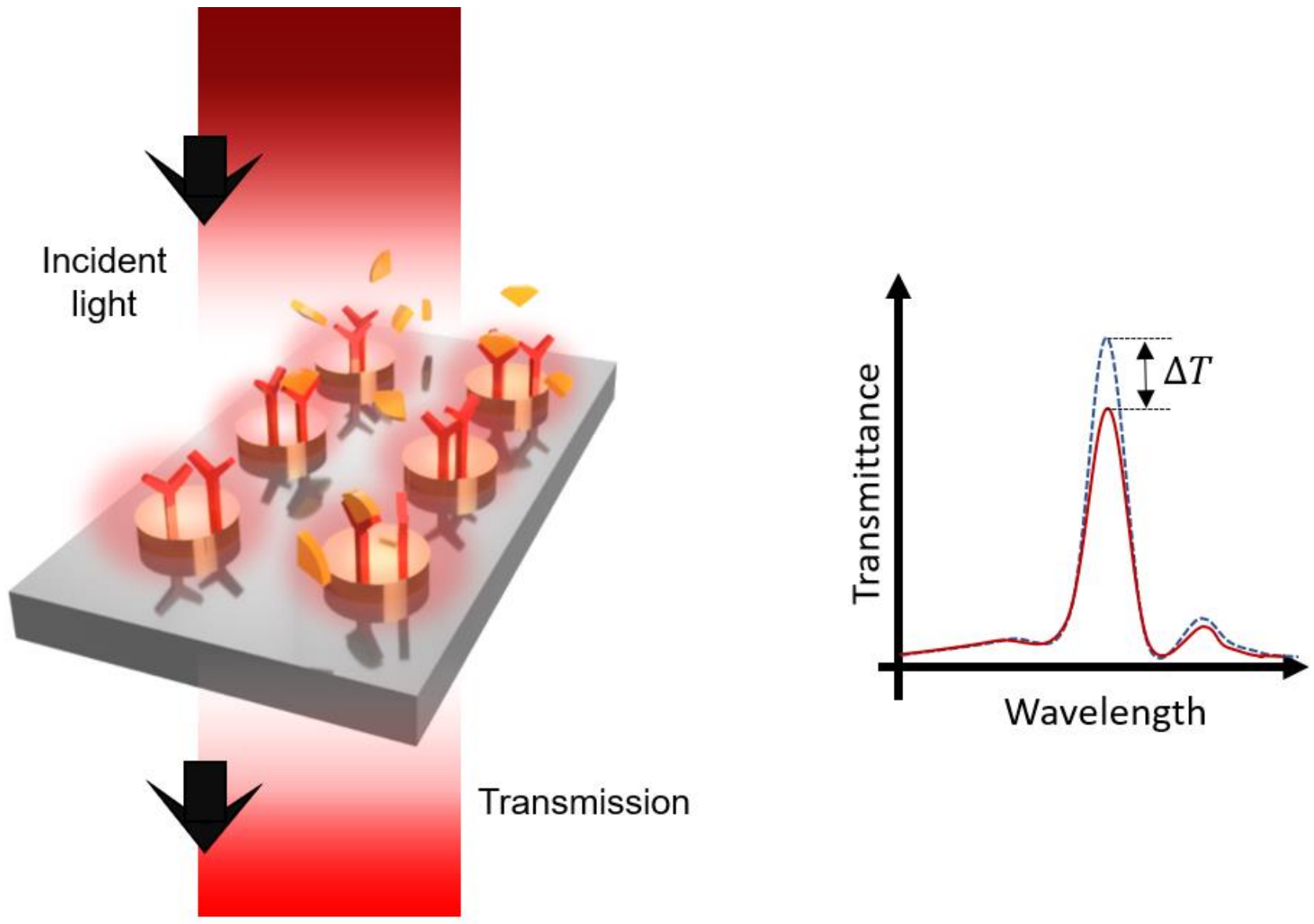

**Fig 14.** The Intensity change of transmitted light based LSPR sensor; typically adapted for mid-infrared absorption spectroscopy.

Another important detecting method using LSPR is surface-enhanced infrared absorption (SEIRA) spectroscopy. Based on the molecular absorption spectroscopy and the fundamental vibrational-rotational transitions of chemical bonds in the wavelength of 3 - 20$\mu m$, mid-infrared (MIR) absorption spectroscopy has been studied vigorously for the label-free detection and identification of molecules in the optical sensor domain. Especially, unlike the near infrared wavelength region, the molecular fingerprint region (700−1500 $cm^{-1}$) in MIR wavelengths contains many absorption bands related to bending and stretching of chemical bonds (such as −C−C−, −C−O−, −C−N−, etc.) that allow the unique identification of biomolecules with high sensitivity and specificity.

Amongst many types of optical based molecular absorption spectroscopy platforms, SEIRA spectroscopy for LSPR devices has been shown its great promise for detecting thin layer of surface-bound nano-molecules due to its tight confinement of surface plasmons on metallic nanostructures, which can significantly enhance the IR absorption of small molecules. Figure 14 shows the typical configuration of SEIRA spectroscopy using LSPR sensor. When the plasmonic resonance peak generated by the metallic nano-antennas is matched with the fundamental vibration signatures of chemical bonds in the biomolecules, the coupling of molecular transitions with the LSPR field on the surface makes significant absorption of corresponding wavelength, so the decrease of transmitted light can be characterized as a sensing result. To obtain more intense IR absorbance, researches have shown various nanostructures including nanorod antenna [135], coaxial-nanogaps[136], and nano cavities[137–139] to enhance the optical confinement and field enhancement of MIR light.

Taking all these device structures and detection methods into account, a number of nanostructure designs have been studied and optimized to apply the plasmonic resonance sensor for the label-free real time detection of various bio-analytes. As shown in Fig. 15a, Lee et al. have shown a multiplex biosensor for cancer biomarkers detection based on the resonance shift of LSPR single gold nanoparticles; the selective sensing results with LOD of 91 fM, 94 fM and 10fM for the α-fetoprotein (AFP), carcino embryonic antigen (CEA) and prostate specific antigen (PSA) analytes are reported by antibody-antigen binding[42]. Amanda et al. reported the detection of Alzhimer disease biomarkers from clinical samples as shown in Fig. 15b; Using surface-confined Ag nanoparticles and sandwich assay, the LOD of < 100 fM for amyloid-â derived diffusible ligands (ADDLs) detection with the specific anti-ADDL antibodies is reported by LSPR induced wavelength shift [109]. To increase the refractive index sensing of LSPR, plasmonic gold mushroom arrays were introduced approaching the theoretical limit of standard SPR configuration with gold surface by Yang et al.; Fig. 15c shows the schematic structure and SEM images of gold mushroom arrays with the detection result of cytochrome c and alpha-fetoprotein, with their LOD down to 200 pM and 15 ng/mL, respectively[140]. Pengyu et al. reported the multiplex serum cytokine analysis by immunoassay enhanced using nano-plasmonic biosensor microarrays (Fig. 15d). Periodically arranged gold nanorod microarray conjugated with corresponding antibodies of each cytokine species (IL-2, IL-4, IL-6, IL-10, IFN-γ, and TNF-R) results in LOD of 5-20 pg/mL from a 1 μL serum sample within 40 min[124]. The hemagglutinin (HA) proteins derived from avian influenza virus detection using SPR device was also reported from Emi et al., with 3.125 nM LOD as shown in Fig. 15e[141]. Recently, Daehan et al. reported the SEIRA spectroscopy using an array of coaxial nano-apertures resonators reported a strong IR absorption enhancement factor of $10^4 \sim 10^5$, and observed ~58 % suppressing of transmitted IR signals from 5nm thick silk protein film as shown in Fig. 15f [136]. Infrared PPA sensor has been reported as in Fig. 15g; Na et al. reported experimental demonstration of Infrared PPA sensor with 99% absorbance at the wavelength of 1.6 $\mu m$, and $FOM_{PPA}$ (eq. (15)) of 87, and 400 nm/RIU sensing sensitivity from glucose solution[133].Intgeration of graphene with SPR sensor as reported by Shuwen et al. have shown an ultrasensitive sensing (Fig. 15i); they reported the LOD of 1 aM for 7.3 kDa 24-mer single-stranded DNA[142]. Moreover, ultrasensitive SPR sensor based on halloysite nanotubes (HNTs)/$MoS_2$/black phosphorous(BP) atomic layers on gold films have been introduced by Guang et al., with the angular and phase detection sensitivities up to $S_A$ = 77.1 $RIU^{-1}$ and $S_P$ = $1.61\times10^5$ $RIU^{-1}$, respectively[143].

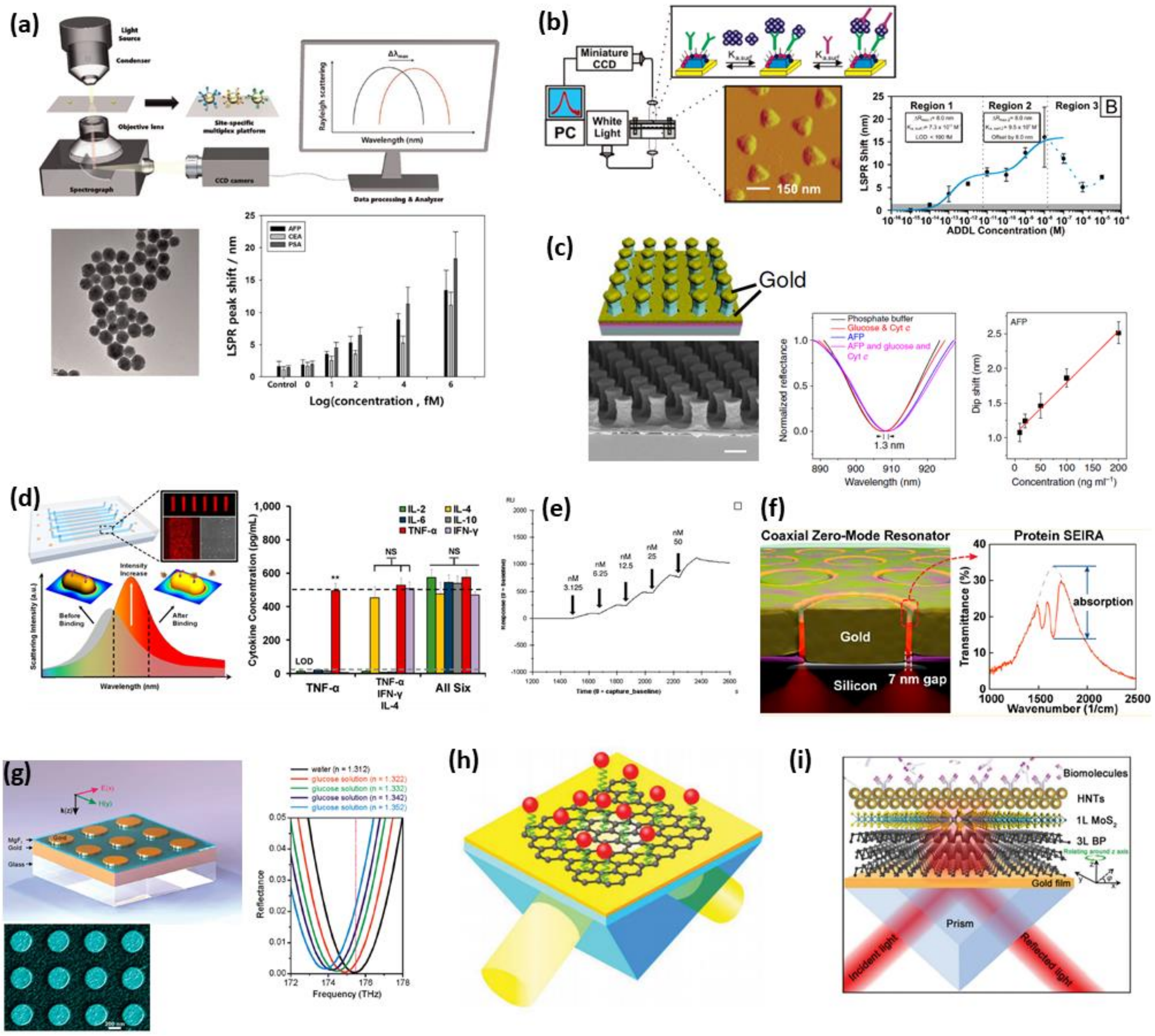

**Fig 15**. Various SPR and LSPR bio sensing applications. (a) Multiplex biosensor for cancer biomarkers detection based on the resonance shift of LSPR single gold nanoparticles; LOD of 91 fM, 94 fM and 10fM for AFP, CEA and PSA, respectively[42]. (b) Alzheimer disease biomarkers from clinical samples using surface-confined Ag nanoparticles and sandwich assay; LOD of < 100 fM for ADDLs[109]. (c) Plasmonic gold mushroom arrays with the LOD down to 200 pM and 15 ng/mL for Cyt c and AFP detecting, respectively[140]. (d) Multiplex serum cytokine immunoassay using gold nanorod microarray conjugated with antibodies to detect cytokine species[124]. (e) Avian influenza virus detection using SPR device with 3.125 nM LOD[141]. (f) Coaxial nano-apertures array resonators for SEIRA spectroscopy with IR absorption enhancement factor of $10^4$~$10^5$, achieving ~58 % suppressing of transmitted IR signals from 5nm thick silk protein film[136]. (g) Infrared PPA sensor with 99% absorbance at the wavelength of 1.6 $\mu m$, FOM of 87, and 400 nm/RIU sensing sensitivity from glucose solution[133]. (h) Ultrasensitive graphene-gold metasurface SPR sensor with LOD of 1 aM for 7.3 kDa 24-mer ssDNA[142]. (i)

Ultrasensitive SPR sensor based on halloysite nanotubes/$MoS_2$/black phosphorous hybrid surface with $S_A$ = 77.1 $RIU^{-1}$ and $S_P = 1.61 \times 10^5$ $RIU^{-1}$. [143]

## *4. SARS-CoV-2 sensing Application*

Here, we review the most up-to-date advances, especially for the coronavirus sensors in plasmonic domain, and introduce well-established plasmonic SARS-CoV-2 biosensing systems. Researchers have demonstrated that using SPR/LSPR-based sensors and corresponding binding biological receptors can effectively and selectively detect coronavirus[103,105,113] (Table 3). Moreover, several researches have already reported SARS-CoV-2 sensing results as in figure 16 [29,126].

**Table 3.** Summary of applications of various plasmonic biosensors for Coronavirus family.

| Analyte | Detection Method | Material | Functionalization | Limit of Detection | Ref |
|---|---|---|---|---|---|
| SARS-CoV-2 | PPT / LSPR | 2D Au nanoislands (AuNIs) | DNA Hybridization | $0.22 \times 10^{-12}$ M | [29] |
| SARS-CoV-2 | SPR | Au film | Antigen-Antibody binding | 100 ng $mL^{-1}$ | [127] |
| HCoV, MERS-CoV | LSPR | An array of carbon electrodes (DEP) modified with gold nanoparticles | Antigen-Antibody binding | HCoV −0.4 pg $mL^{-1}$, MERS-CoV −1.0 pg $mL^{-1}$ | [110] |
| SARS-CoV | ELISA / LSPR | PMMA optical fiber / AuNPs | Antigen-Antibody binding | 1 pg $mL^{-1}$ | [111] |
| Multiple Respiratory Viruses (Influenza A, Influenza B, H1N1, RSV, PIV 1,2,3, Adenovirus, SARS-CoV) | SPR | Au film / glass wafers | DNA Hybridization | Influ A – 5 nM<br>Influ B – 1 nM<br>PIV 1 – 1 nM<br>PIV 2 – 2.5 nM<br>PIV 3 – 3.5 nM<br>RSV – 3 nM<br>ADV – 0.5 nM<br>SARS – 2 nM<br>H1N1 – 3 nM | [113] |

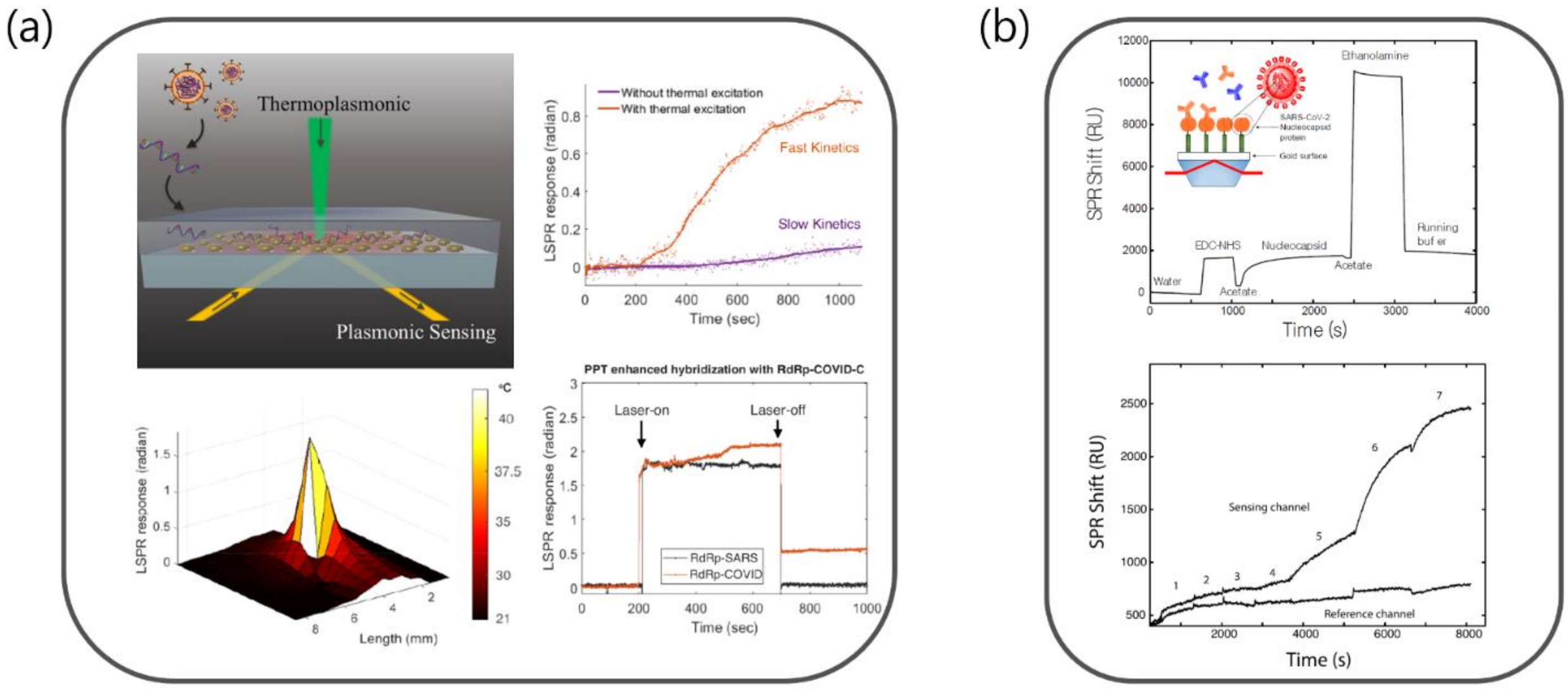

**Fig 16.** State-of-the-art plasmonic biosensor for SARS-CoV-2 sensing applications. **(a)** Photothermal enhanced LSPR biosensor for nucleic acid sequences detection from SARS-CoV-2. The schematic shows the configuration of LSPR device consists of gold nanoparticles where the local heat is generated by the thermo-plasmonic effect. The graph shows the sensing enhancement by the thermal excitation with the limit of detection of 0.22 pM and the sensing selectivity between the RdRp-SARS and RdRp-COVID hybridization; Reprinted from[29]. **(b)** SARS-CoV-2 antibodies sensing with SPR in undiluted human serum. The schematic shows the SPR device

configuration, coated with a peptide monolayer to detect SARS-CoV-2 antibodies, and the graphs show the sensor-gram and SPR sensing results with the limit of detection value of 100 ng $mL^{-1}$; Reprinted from[126].

A recent article reported the dual-functional plasmonic photothermal biosensors for SARS-CoV-2 detection[29]. The authors demonstrated a highly sensitive, fast, and reliable SARS-CoV-2 virus detection capability by integrating the plasmonic photothermal (PPT) effect and conventional LSPR sensing transduction on a single gold nanoislands (AuNI) chip. The two-dimensional AuNIs functionalized with cDNA receptors (RdRp-COVID-C) can perform a selective detection of the RdRp-COVID through DNA hybridization, and the LOD down to the concentration of 0.22 pM is reported. Lately, Djaileb et al reported the antibody detection specific against SARS-CoV-2 in undiluted human serum using SPR sensing[126]. A SPR sensor coated with a peptide monolayer detected anti-SARS-CoV-2 antibodies in the nanomolar range (LOD of 100 ng $mL^{-1}$) was performed and analyzed within 15 minutes.

Furthermore, several researches reported that the sensitivity and the signal-to-noise ratio (SNR) of conventional enzyme-linked immunosorbent assay (ELISA) or fluorescence-linked immunosorbent assays (FLISA) tests can be significantly improved by applying the 'add-on' plasmonic particles without altering their workflow[50,103]. As the ELISA test is widely used for precise SARS-CoV-2 detection[103], plasmon enhanced ELISA/FLISA tests can be applied to COVID-19 sensing as well.

Although aforementioned SARS-CoV-2 sensing applications[29,126] have shown great performances of sensitive and selective sensing, a huge potential for more sensitive, accurate and fast on-chip sensing with less complex system is still remained in LSPR biosensor domain. For example, the sensing systems in Fig. 16[29,126] require the prism coupler to couple the incident light into SPR device with an accurate incident angle. It requires very sensitive alignment of optical devices which makes the overall system complex and hard to be integrated with sources and detectors. However, as described earlier in Table. 2, the incident light can be coupled into LSPR sensors directly without the external couplers, and this normal-incident angle can make the alignment easier, in turn mitigate the complexity of the system and make the possibility of fully integrated on-chip sensing; moreover, due to the capability of sensor miniaturization through LSPR nanostructures, label-free, real-time, and parallel detection with multiple channels with high-specificity are achievable. Furthermore, improving the sensitivity by applying advanced materials has incited a great interest for various optical biosensor applications. For plasmonic biosensors, the ultrasensitive graphene and 2D material enhanced SPR devices have been reported as shown in Fig. 16 (h) and (i)[142,143], and the experimental sensing result with LOD value approaching 1 atto M has been shown[142,144]. Accordingly, the advanced material enhanced LSPR biosensors are anticipated to enable the possibility of highly sensitive, accurate and fast point of care lab-on-a-chip integrated sensor with unprecedented high sensitivity. The detailed discussion of emerging nanomaterials for optical biosensors are described in section V.

## IV. SARS-COV-2 BIOSENSOR: DESIGN AND IMPLEMENTATION

To develop an accurate estimate of COVID-19 biosensing functioning mechanism, a simulation model need to be first designed. Here, in a proposed simulation model, COVID-19 is approximated to be a solid sphere core containing RNA covered with a membrane protein with radiuses of $r_1$ and $r_2$, respectively (Fig. 17a)[45]. Thus, the effective RI of the virus is calculated by taking a volume weighted sum of the two refractive indices:

$$n_{eff} = \frac{n_1 V_1}{V_1 + V_2} + \frac{n_2 V_2}{V_1 + V_2} = \frac{n_1 + n_2(\eta^3 - 1)}{\eta^3},\ r_2 = \eta r_1, \quad \text{(x)}$$

where $n_1$ ($V_1$) and $n_2$ ($V_2$) are the total RI of the RNA and the membrane protein volume, respectively. As the RI of the virus is determined mainly by material composition rather than its geometrical size, $\eta$ is a constant value for the same kind of virions. ($\eta_{COVID\text{-}19}$=1.25 average value of several measurements of transmission electron microscopy (TEM) pictures[145]). We use the SWGR design to simultaneously take advantage of the enhanced binding surface and strong light-substance interaction. As shown in Fig. 17 c and e, the energy mode is distributed between the gratings as well. In order to further improve the SWG waveguide functioning in the sub-wavelength range, the grating period Λ, the waveguide width w, and the fill factor are designed to be 230nm, 1.23µm, and 0.5µm, respectively. For the SWGR, the radius $R$ is set as 5µm with the corresponding FSR[57] of 25nm at 1550nm. Here simulation system includes a 220nm-Silicon top-layer with a 3µm buried oxide (BOX) wafer and a liquid solution, with the refractive index of $n_{clad}$ to be 1.35[31]. Adopting our previous designs features[38,39], we optimized a high-Q SWGRs by utilizing a trapezoidal (T) silicon pillars and reducing bending loss by ~50% compared to a conventional rectangular silicon pillar. We therefore set the SW waveguide width to be 0.5µm (correlated to the fundamental mode of transverse electric (TE)) and studied the effect of the trapezoidal width. It is noted that to obtain the lowest bending loss of the T-SWG waveguide, we employ the particle swarm method for the optimization process. Three parameters (w, $A_1$, $A_2$) are optimized and are defined as the width, the tuning factor of the outer and inner filling factor of the SWG, respectively (as shown in the inset figure of Fig. 17b). Considering the limitations of the design for fabrication, the slot between gratings are pre-set to be larger than 60nm, thus $A_1$ and A2 are limited to be (1, 2) and (0, (1-60nm/$\Lambda$)/$f$), respectively. At the same time, to keep the SWG working in subwavelength regime ($\Lambda << \boldsymbol{\lambda}/2n_{eff}$), $\Lambda$ is safely set to be 230nm and $f$ is simply set to be 0.5 with no optimization. Furthermore, to make the SWG waveguide work as a single or few mode waveguide, the width of the gratings is set less than 2μm. All in all, the ranges of w, $A_1$, and $A_2$ are set to be (0.5μm, 2μm), (1,2), and (0, 0.522), respectively. The FOM is defined to achieve the lowest bending loss with the bend radius of 5μm. Based on the optimization measures taken, we finally achieved a bending loss as low as 0.0279 dB/cm with the optimized (w, $A_1$, $A_2$) = (1.23μm, 1, 0.522). By adjusting the coupling gap between the insertion SWG waveguide and the designed SWGR, the Q can be as high as ~50000 (the resonance at 1557.6nm) with a broad FSR of 25 nm, as shown in Fig. 17a. We also optimized the 10μm radius SWGR (not shown in the Fig.17f, achieving a loaded Q of ~75000 (the resonance at 1552.1nm) with the FSR of 11nm with (w, $A_1$, $A_2$) = (1.23μm, 1, 0.522), at the same waveguide-ring cross-coupling coefficients. Needless to say as quality factor of the ring becomes higher, the fabrication tends to be more challenging. Bulk RI sensitivity (shown in the Fig.11f inset figure) in the buffer solution is calculated to be $S_{res} = \frac{\Delta\lambda_{res}}{\Delta n_{clad}} = 400\ nm/RIU$. Thus the *iDL* can be calculated as low as ~7.5e$^{-5}$ RIU. Note that *iDL* performances can be further improved by exploiting a larger radius ring or by further achieving the critical coupling condition given the predictable higher Q, while making trade-off between the performance and the sensor size or the resonance peak extinction ratio.

**Surface sensing:** To evaluate the specific sensing ability of the proposed device for COVID-19, surface sensing performances are analysed by considering the device immersed in buffer solution, bonded by several surface layers (generated in the sensing preparation process) including the ~2-3nm surface oxide layer, ~10nm functionalization layer and bonded antibody (protein layers), and the bonded virus particles layer in detection process. In simulations, the preparation process generated layers are further simplified to be a uniform layer (RI: 1.45) with a thickness of 15nm, and the bonded virus layer is simplified as a uniform layer with a thickness of 125nm (the maximum diameter of the COVID-19 virus) (Fig. 17a). It is noted that the equivalent RI of the virus layer ($n_{binding}$) depends on the number of bonded virus, which is a function of the virus concentration and binding processing time, and is dominated by the concentration in real sensing process with a given binding time. Thus, the SWGR sensing performance can be evaluated by calculating the $n_{binding}$ response of the device, with the $n_{binding}$ ranging from 1.35 (no binding) to 1.5 (full binding). Simulation results in Fig.17f shows the functionalization and the full binding process induces a shift of 3.41nm and 1.14, respectively. The obvious simultaneous measurable shifts in the FSR range ($\Delta\lambda_{res} < FSR$) and experimental values ($\Delta\lambda_{res} \gg 1pm$) indicates the promising potentials of the proposed device in detecting the COVID-19 virus or simply being as a chemical/bio-sensor in future practical applications.

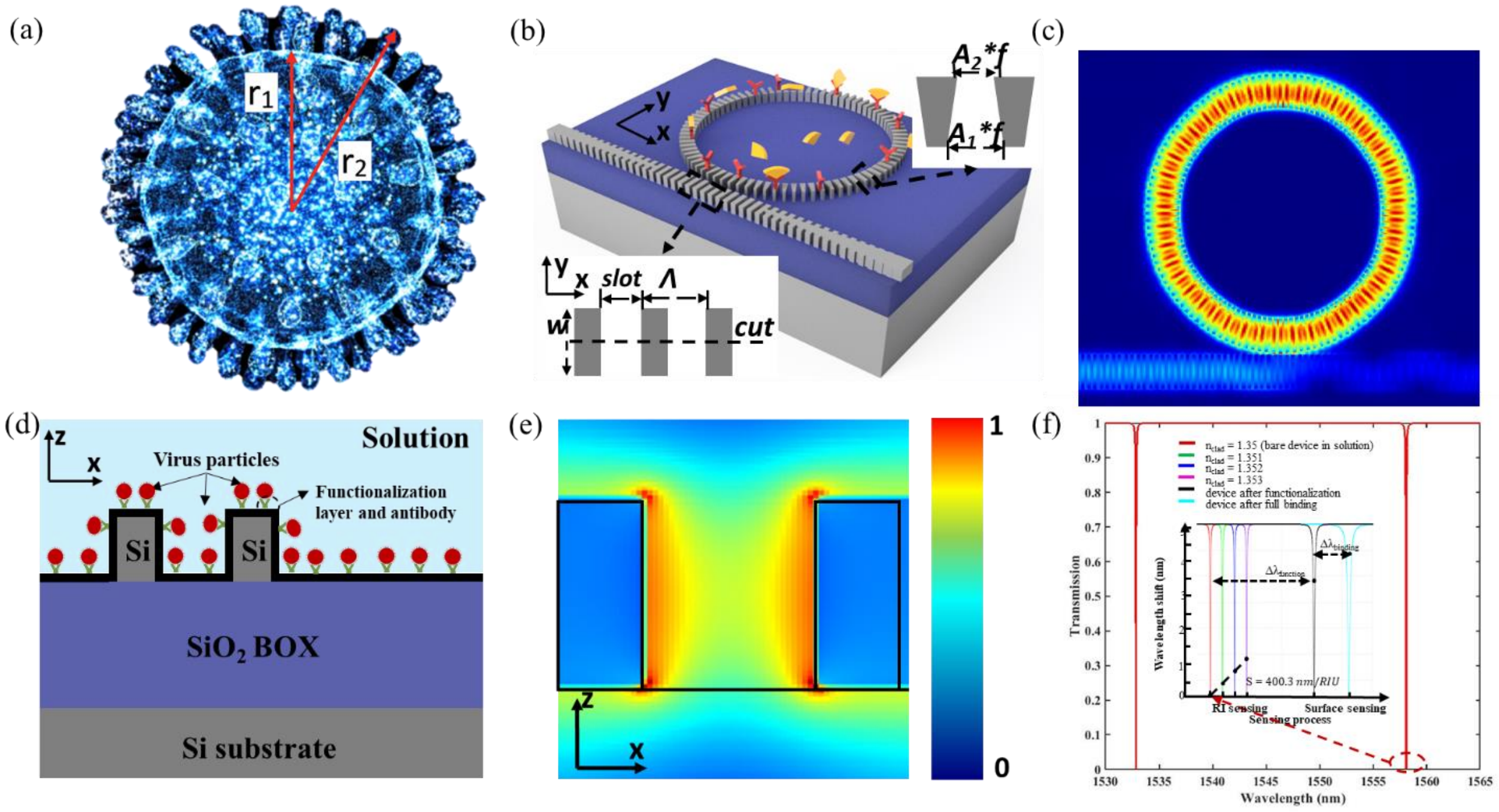

Fig. **17** (a) Schematic the COVID-19 virus, (b). Schematic illustration of the SWGR biosensor with a low loss in/out coupling via lineally tapering the input and the output gratings, (c) Top view of the simulated SWGR biosensor for a fundamental TE mode at 1550nm, (d) Details of the bonded COVID-19 on the substrate, (e) Simulated intensity distribution of the electric field at 1550nm. Light penetrates between the slot waveguides providing better light matter interaction with the COVID-19 analyte, and (f) Calculated transmission spectrum of the proposed device. Inset figure shows the calculated bulk RI sensing and surface sensing responses of the device, with the bulk RI sensing shows a sensitivity of 400nm/RIU, and surface sensing shows the total wavelength shifts of 3.41nm nm and 1.14 nm after the functionalization process and the full binding process, respectively.

## V. TWO-DIMENSIONAL MATERIALS OPTICAL BIOSENSORS

Latest major advancements in preparation, development, and utilization of new low-dimensional materials has been attractive for development of modern miniaturized biosensors and immunosensors. Graphene and its analogous 2D materials such as transition metal dichalcogenides (TMDCs), carbides and nitrides(Mxenes), hexagonal boron nitride (h-BN), black phosphorus (BP) and transition metal oxides (TMOs) have attracted great attention to be used as transducer due to combined high sensitivity and selectivity for biosensors.

### *A. Graphene and graphene oxide (Gr and GO)*

Graphene has been regarded a revolutionary material ever-since its first introduction in 2004[146], given its extraordinary optical and electronic properties [32,147–161]. Since then, graphene has also shown an immense potential in different applications and a great deal of graphene based biomolecular sensors have been specifically developed by paying especial attention to its biocompatibility and high specific surface area[45,162–164] . On top of that, unique and ideal optical properties such as broadband and tunable absorption and polarization-dependent nonlinear optical effects make graphene a promising candidate to be employed for optical based biosensors (Fig 18). The introduction of advanced biosensors through graphene electrical and optical qualities in general has delivered extraordinary sensitivity, detection level, resolution and response time in many devices (Fig 18. c and d)[32,106,151,152,154,163–167]. Point-of-Care biosensors has also shown enhanced sensitivity in graphene based electrochemical biosensors as well[168]. While pristine graphene has seen applications in biosensors devices, its derivative GO has been subjected to a wealth of investigation for rapid detection, disinfection of pathogens and enzyme assays, making it a key material for a variety of biomedical applications. Graphene oxide has been a suitable precursor for graphene and its biosensors applications especially due to its attractive distinctive properties like good water dispersibility, facile surface modification and to be more specific photo-luminescence for optical biosensors. Jin et al demonstrated a functionalized graphene oxide wrapped around SiO2 which possess superior RNA sensitivity and limit of detection up to 1 fM with the potential to show even higher sensitivity values[169]. They showed the high electron conduction and higher surface area in spherical morphology has been specially effective to improve the sensitivity and limit of detection[169]. Graphene unique electrical properties has also been exploited effectively to develop different transistor based label-free biosensors including COVID-19 detection system[39] (fig 18.a). Aside from the field-effect-transistor-based graphene biosensor, which relies mainly on current changes, providing easier mass-scale production with satisfying sensitivity, it’s limited sensing capability along with being damaging to living cells make its application limited compared to analogous optical ones[170] [Table 3].

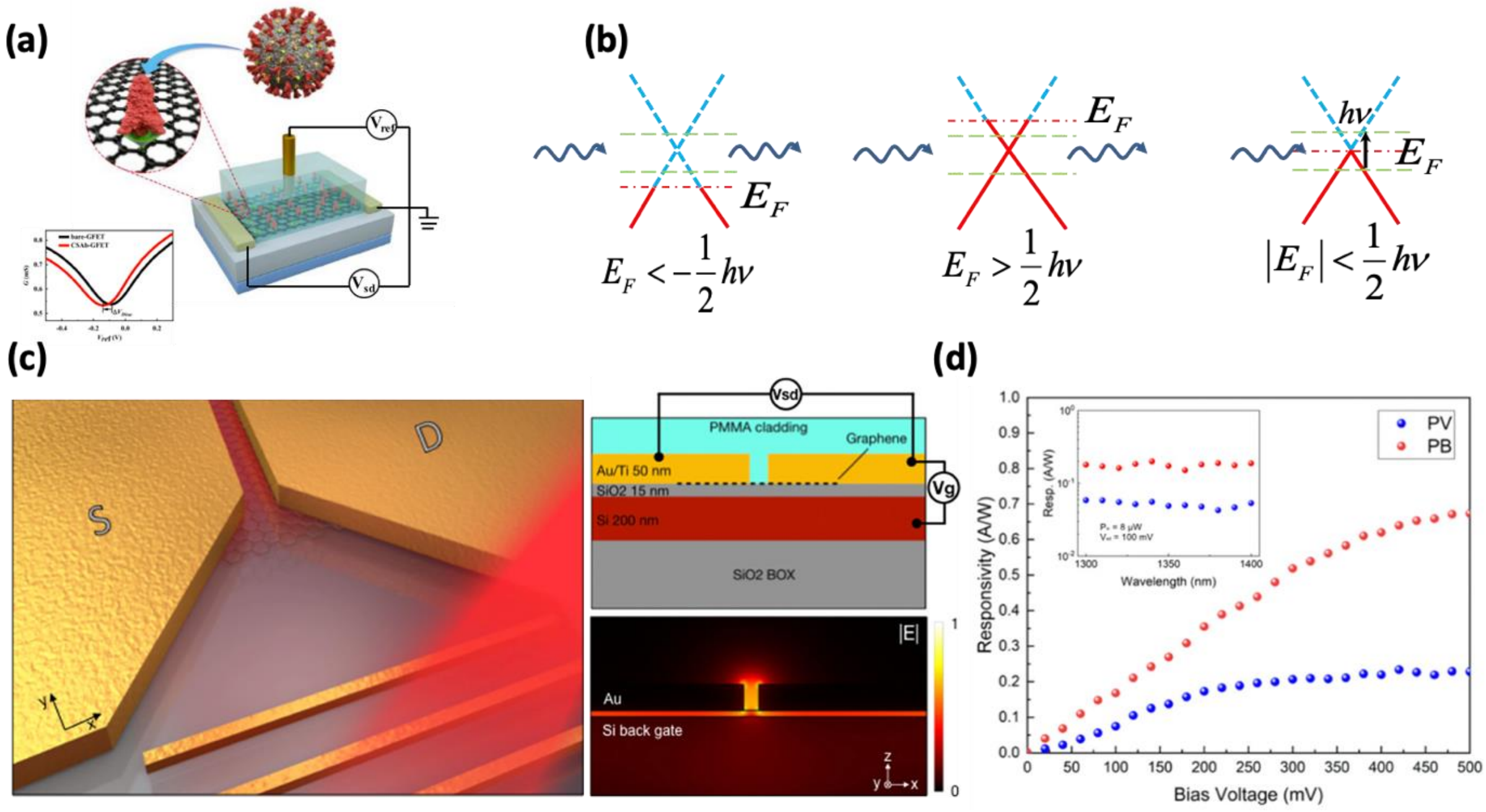

**Fig 18**. a) Schematic of functionalized graphene FET immunosensor. Inset shows the effect of functionalization with CSAb vs non-functionalized graphene on increasing sensitivity to COVID-19 antibodies spike protein[39]. b) The band structure of graphene under different gate voltages of positive, almost zero and negative from left to right, respectively. (c) (left) Three-dimensional representation of a graphene photodetector[154] (right) Cross-sectional schematic of the device, where the Ti/Au metallic structures are in close proximity to each other for forming the plasmonic slot waveguide d) Measured responsivity vs bias voltage of graphene photodetector. Inset shows a broadband responsivity from 1300 to 1400 nm. The spectral response of the graphene-slot detector is mainly limited by the relatively narrow metallic grating coupler operating bandwidth.

The interaction of graphene with light is a determining factor in effectiveness of integrated graphene with different photonic and optoelectronic devices including biosensors; a high-quality graphene-based biosensor requires a strong interaction between graphene and light. An undoped graphene monolayer has a constant broadband absorption equal to %2.3[171]. It has been shown that the Fermi level of graphene can be tuned by either electrical or chemical dopant beyond half of the photon energy[156,172,173] (Fig 18) which is critical for sensitivity enhancement in a graphene based label-free biosensor. This happens as a result of graphene becoming transparent due to Pauli blocking effect. This in turn leads to significant refractive index change (RI) and phase shift of the input light. High nonlinearity in graphene opens the door to a broad range of graphene based nonlinear plasmonic sensor applications[174].

**Table 3.** The difference between a graphene-based electrical and optical sensor [adapted[32,155,156]].

| Sensor type | Working principle | Advantages | Disadvantages |
|---|---|---|---|
| **Graphene electrical sensor** | Ambipolar behaviour of the graphene and its doping by applying gate voltage enables sensing by induced changes in drain-source conductivity of the graphene channel upon the binding of the sample to the receptor-functionalized graphene | Small size, fast electron transfer and response time, high sensitivity and reduced surface contamination | Limited sensing capacity (only current changes), low spatial resolution, damaged testing sample |
| **Graphene optical sensor** | Enhanced polarization absorption and broadband absorption under total reflection enables sensing by using attenuated total reflection to detect refractive index changes near the surface | High spatial resolution, wide detection range, high sensitivity and precision, accurate and fast detection, unlabeled samples | Photocurrent is too small due to low absorption rate of graphene |

### *B. Graphene Surface Plasmon Resonance*

Researchers have investigated plentiful ways of enhancing the sensitivity of the SPR sensor, including but not limited to the use of resonant structures such as metal nanoparticles and optical gratings. As shown before, in a typical SPR or LSPR biosensor, a thin metallic film like Ag or Au gets deposited on prism to separate it from the sensing area. The deposited metallic film brings out the propagation of surface plasmon at visible light frequency. Gold is preferred as it provides better resistance to oxidation and corrosion in different environments. The intrinsic defects in gold and silver-based biosensors like oxidization of metal, poor adsorption to biomolecules and hence limited sensitivity and accuracy leads researchers to seek methods of alternatives. In view of the defects of biosensors based on gold and silver film, graphene-based biosensor has been developed. Graphene provides a highly sensitive non-oxidizing receptor substrate to analytes[152,166,175]. Moreover, graphene also helps to adsorb biomolecules better, because of π-π stacking, which increases the system's affinity for these molecules (Fig 19)

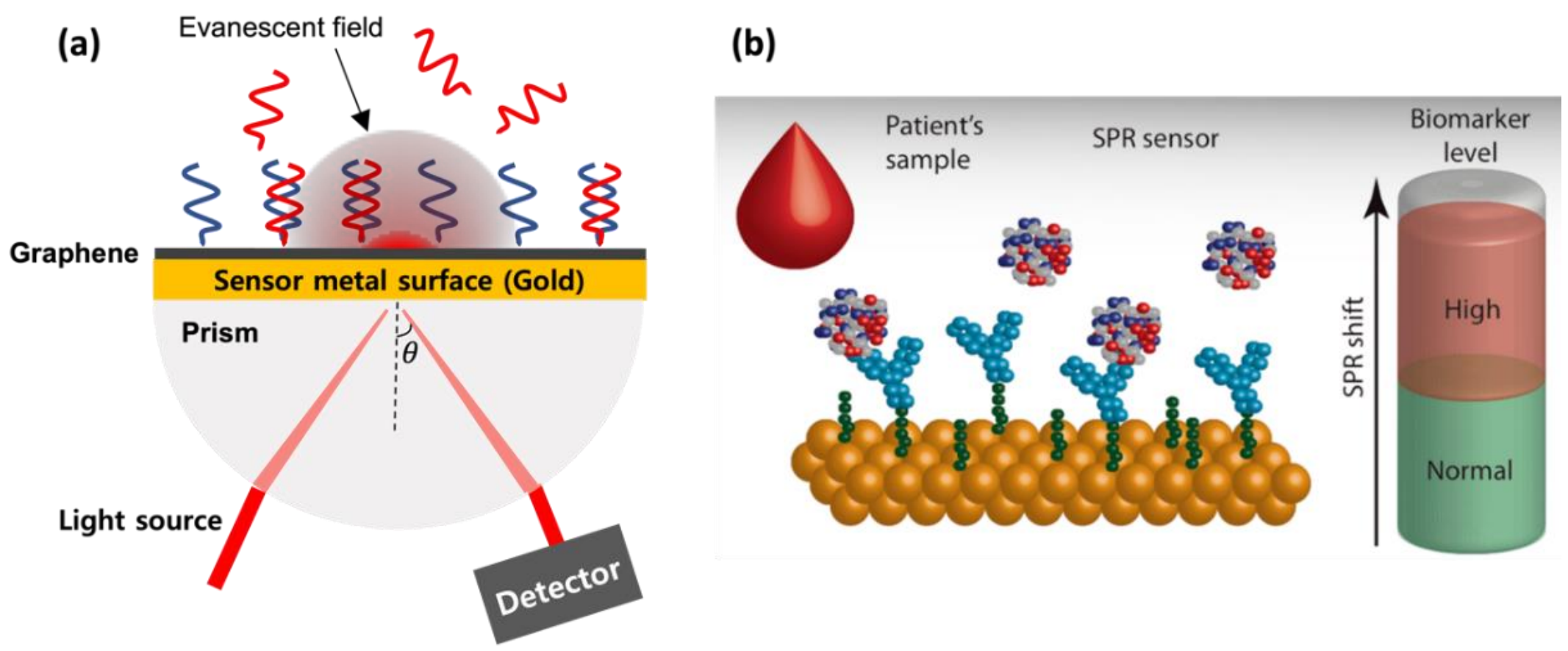

**Fig 19**. a) Schematic of the surface plasmon resonance(SPR) enhanced with graphene layer [adapted[45]] and b) Schematic of a SPR biosensor functioning mechanism(Reprinted[44]).

The graphene plasmonic nano-islands can demonstrate nonlinearities two orders of magnitude higher than the their non-graphene counterparts of equal size[174]. These nano-islands can be used as nonlinear plasmonic biosensors, as the presence of an individual molecule is sufficient to trigger a large change in the nonlinear response of the graphene plasmons. The effect of functionalization on the sensitivity enhancement of graphene and using it especially for biosensors have been discussed in detail[29,176–179].

*C. Transition metal dichalcogenides (TMDCs)*

Transition metal dichalcogenides(TMDs) are a class of materials including but not limited to MoS2, MoTe2 and TiNb. Bulk crystals of TMDs have been long known to researchers throughout the world but the new discoveries in isolation of their two-dimensional(2D) structures has led to a plethora of new properties Unique physical and chemical properties of 2D TMDs including high surface to volume ratio, sizable bandgap and high absorption coefficient along with high potential for redox reactions on its surface makes them capable of superior molecular sensitivity[180].
MoS2 has recently found widespread applications as a fluorescence probe in the detection of various biological and environmental analytes. As discussed in the previous section, MoS2 possesses good optical properties like fluorescence and have been utilized to design various sensing devices. The possession of PL characteristics provides the possibility of using such structures in fluorescence-based applications, which can be used to trace, image, and sense biological components. The Raman characteristics of the MoS2 monolayer is also a function of the dimension and permittivity of the environment. Attaching biological components may alter such characteristics and thus can be used as a biosensing principle.

## VI. CONCLUSION

Emerging pandemics and epidemic diseases like COVID-19 brings out a high demand in advancement and research in medical detection and treatment methods. The optical biosensors provide a fast detection (< 1min) of such a virus at very low concentrations (~1 fM). However, they need to be designed and functionalized to be the most absorptive to the target analyte. The ideal label-free biosensor is cheap, disposable, or reusable, compact, and semi-automatic. Although most efforts in biosensors have been focused on protein biomarkers, other targets such as small molecules and nucleic acids are crucial in expanding the application of biosensors including optical ones. A common challenge for optical biosensors is to reach the capability of performing the measurement in real complex samples, avoiding, or limiting the sample preparation phase. Developing label-free biosensor is aligned with that purpose as the need for on-site detection techniques is boosting as the world post-COVID-19 pandemic will never be like before.

**Acknowledgment**

This research team is supported by AFOSR MURI silicon nanomembrane research center (FA9550-08-0394), NIH ( HHSN261201200043C), US ARO ( W81XWH-14-C-0029), NASA STTR phase I program (80NSSC18P2146), AFOSR STTR phase I program (FA9550-19-P-0004), AFOSR SBIR phase II program (FA9550-19-C-0003) and the State of Texas. A.A. is supported by UT Austin RAship and C.W. is supported by scholarship funding (# 201806470008) for his basic PhD research in UT Austin.

**Data Availability**

The data that support the findings of this study are available from the corresponding author upon reasonable request.

**References**

[1] D. Wrapp, N. Wang, K.S. Corbett, J.A. Goldsmith, C. Hsieh, O. Abiona, B.S. Graham, and J.S. Mclellan, Science **1263**, 1260 (2020).

[2] C. Hsieh, J.A. Goldsmith, J.M. Schaub, A.M. Divenere, H. Kuo, K. Javanmardi, K.C. Le, D. Wrapp, A.G. Lee, Y. Liu, C. Chou, P.O. Byrne, C.K. Hjorth, N. V Johnson, A.W. Nguyen, J. Park, N. Wang, D. Amengor, J.A. Maynard, I.J. Finkelstein, and J.S. Mclellan, BioRxiv (2020).

[3] C. Weiss, M. Carriere, L. Fusco, I. Capua, J.A. Regla-nava, M. Pasquali, J.A. Scott, F. Vitale, M.A. Unal, C. Mattevi, D. Bedognetti, and A. Merkoc, ACS Nano **14**, 6383 (2020).

[4] L. Premkumar, S. Immunol, L. Premkumar, B. Segovia-chumbez, R. Jadi, D.R. Martinez, R. Raut, A. Markmann, C. Cornaby, L. Bartelt, S. Weiss, Y. Park, C.E. Edwards, E.M. Scherer, N. Rouphael, S. Edupuganti, D. Weiskopf, L. V Tse, Y.J. Hou, D. Margolis, A. Sette, M.H. Collins, J. Schmitz, R.S. Baric, and A.M. De Silva, Sci. Immunol. **8413**, 1

(2020).

[5] L. Du, Y. He, Y. Zhou, S. Liu, and B.J. Zheng, Nature Microbiology **7**, 226 (2009).

[6] T. Kuiken, R.A.M. Fouchier, M. Schutten, G.F. Rimmelzwaan, G. Van Amerongen, D. Van Riel, J.D. Laman, T. De Jong, G. Van Doornum, W. Lim, A.E. Ling, P.K.S. Chan, J.S. Tam, M.C. Zambon, R. Gopal, C. Drosten, S. Van Der Werf, N. Escriou, J. Manuguerra, K. Stöhr, and J.S.M. Peiris, The Lancet **362**, 263 (2003).

[7] W. Li, M.J. Moore, N. Vasilieva, and J. Sui, Nature **426**, 450 (2003).

[8] M.A. Marra, S.J.M. Jones, C.R. Astell, R.A. Holt, A. Brooks-wilson, Y.S.N. Butterfield, J. Khattra, J.K. Asano, S.A. Barber, S.Y. Chan, A. Cloutier, S.M. Coughlin, D. Freeman, N. Girn, O.L. Griffith, S.R. Leach, M. Mayo, H. Mcdonald, S.B. Montgomery, P.K. Pandoh, A.S. Petrescu, A.G. Robertson, J.E. Schein, A. Siddiqui, D.E. Smailus, J.M. Stott, G.S. Yang, F. Plummer, A. Andonov, H. Artsob, N. Bastien, K. Bernard, T.F. Booth, D. Bowness, M. Czub, M. Drebot, L. Fernando, R. Flick, M. Gray, A. Grolla, S. Jones, H. Feldmann, A. Meyers, A. Kabani, Y. Li, S. Normand, U. Stroher, G.A. Tipples, S. Tyler, R. Vogrig, D. Ward, B. Watson, R.C. Brunham, M. Krajden, M. Petric, D.M. Skowronski, C. Upton, and R.L. Roper, Science **300**, 1399 (2003).

[9] J.S.M. Peiris, S.T. Lai, L.L.M. Poon, Y. Guan, L.Y.C. Yam, W. Lim, J. Nicholls, W.K.S. Yee, W.W. Yan, M.T. Cheung, V.C.C. Cheng, K.H. Chan, D.N.C. Tsang, R.W.H. Yung, T.K. Ng, and K.Y. Yuen, The Lancet **361**, 1319 (2003).

[10] P.A. Rota, M.S. Oberste, S.S. Monroe, W.A. Nix, R. Campagnoli, J.P. Icenogle, B. Bankamp, K. Maher, M. Chen, S. Tong, A. Tamin, L. Lowe, M. Frace, J.L. Derisi, Q. Chen, D. Wang, D.D. Erdman, T.C.T. Peret, C. Burns, T.G. Ksiazek, P.E. Rollin, A. Sanchez, S. Liffick, B. Holloway, J. Limor, K. Mccaustland, M. Olsen-rasmussen, R. Fouchier, A.D.M.E. Osterhaus, C. Drosten, M.A. Pallansch, L.J. Anderson, and W.J. Bellini, Science **300**, 1394 (2003).

[11] F. Li, Annual Review of Virology **3**, 237 (2016).

[12] S. Jiang, C. Hillyer, and L. Du, Trends in Immunology **41**, 355 (2020).

[13] A.Z. Wec, A.Z. Wec, D. Wrapp, A.S. Herbert, D.P. Maurer, D. Haslwanter, M. Sakharkar, R.K. Jangra, M.E. Dieterle, A. Lilov, D. Huang, L. V Tse, N. V Johnson, C. Hsieh, N. Wang, J.H. Nett, E. Champney, I. Burnina, M. Brown, S. Lin, M. Sinclair, C. Johnson, S. Pudi, R.B. Iii, A.S. Wirchnianski, E. Laudermilch, C. Florez, J.M. Fels, C.M.O. Brien, B.S. Graham, D. Nemazee, D.R. Burton, R.S. Baric, J.E. Voss, K. Chandran, J.M. Dye, J.S. Mclellan, and L.M. Walker, Science **7424**, 1 (2020).

[14] F. Li, Journal of Virology **89**, 1954 (2015).

[15] H. Hofmann, K. Pyrc, L. Van Der Hoek, M. Geier, B. Berkhout, and S. Po, PNAS **102**, 7988 (2005).

[16] Z. Qinfen, C. Jinming, H. Xiaojun, Z. Huanying, H. Jicheng, F. Ling, K. Li, and Z. Jingqiang, Journal of Medical Virology **337**, 332 (2004).

[17] Y. Guo, C. Korteweg, M.A. Mcnutt, and J. Gu, Virus Research **133**, 4 (2008).

[18] Y. Ding, L. He, Q. Zhang, Z. Huang, X. Che, J. Hou, H. Wang, H. Shen, L. Qiu, L. Zhuguo, J. Geng, Cai, Junjie, H. Huixia, X. Li, W. Kang, W. Desheng, P. Liang, and S. Jiang, J Pathol **203**, 622 (2004).

[19] R. Syndrome, 1136 (2007).

[20] J.F. Zhong, L.P. Weiner, K. Burke, and C.R. Taylor, Journal of Virological Methods **144**, 98 (2007).

[21] B. Kaltenboeck and C. Wang, Advances in Clinical Chemistry **40**, 219 (2020).

[22] T.K. Nagasse-sugahara, J.J. Kisielius, M. Ueda-ito, S.P. Curti, C.A. Figueiredo, and Á.S. Cruz, Rev. Inst. Med. Trop. S. Paulo **46**, 315 (2004).

[23] M.G. De Castro, R. Maria, R. Nogueira, H.G. Schatzmayr, M.P. Miagostovich, and R. Lourenço-de-oliveira, Mem Inst Oswaldo Cruz, Rio de Janeiro **99**, 809 (2004).

[24] S.E.J. Gibbs, A.E. Ellis, D.G. Mead, A.B. Allison, J.K. Moulton, E.W. Howerth, and D.E. Stallknecht, Journal of Wildlife Diseases **41**, 354 (2005).

[25] G. Seo, G. Lee, M.J. Kim, S. Baek, M. Choi, K.B. Ku, C. Lee, S. Jun, D. Park, H.G. Kim, S.S. Il Kim, J. Lee, B.T. Kim, E.C. Park, and S.S. Il Kim, ACS Nano **14**, 5135 (2020).

[26] L.E. Lamb, S.N. Bartolone, E. Ward, and M.B. Chancellor, PLoS ONE **15**, 6 (2020).

[27] K.K. To, O. Tak-Yin Tsang, W. Leung, A.R. Tam, T. Wu, D.C. Lung, C.C. Yip, J. Cai, J.M. Chan, T.S. Chik, D.P. Lau, C.Y. Choi, L. Chen, W. Chan, K. Chan, J.D. Ip, A.C. Ng, R.W. Poon, C. Luo, V.C. Cheng, J.F. Chan, I.F. Hung, Z. Chen, H. Chen, K. Yuen, F. Richard, C. Yu, M. Tam, M. Mei, T. Shaw, F. Hong, M. Tong, and M. Lee, The Lancet Infectious Diseases **20**, 565 (2020).

[28] S. Hu, S. Qiao, J. Pan, B. Kang, J. Xu, and H. Chen, Talanta **179**, 9 (2018).

[29] G. Qiu, Z. Gai, Y. Tao, J. Schmitt, G.A. Kullak-ublick, and J. Wang, ACS Nano **14**, 5135 (2020).

[30] M.S. Cheng and C.-S. Toh, Analyst **138**, 6219 (2019).

[31] Homola, Jiri, S.S. Yee, and G. Gauglitz, Sensors and Actuators B **54**, 3 (1999).

[32] Z. Li, W. Zhang, and F. Xing, International Journal of Molecular Sciences Review **20**, 24 (2019).

[33] V. M. N. Passaro, F. Dell'Olio, B. Casamassima, and F. De Leonardis, Sensors **7**, 508 (2007).

[34] C.S. Huertas, O. Calvo-lozano, A. Mitchell, and L.M. Lechuga, Frontiers in Chemistry **7**, 1 (2019).

[35] E. Luan, H. Shoman, D.M. Ratner, K.C. Cheung, and L. Chrostowski, Sensors **18**, 3519 (2018).

[36] J. Chen, L. Wu, J. Zhang, L. Zhang, D. Gong, Y. Zhao, S. Hu, Y. Wang, X. Hu, B. Zheng, K. Zhang, H. Wu, Z. Dong, Y. Xu, Y. Zhu, X. Chen, L. Yu, and H. Yu, MedRxiv (2020).

[37] Z. Alom, M.M.S. Rahman, M.S. Nasrin, T.M. Taha, and V.K. Asari, ArXiv (2020).

[38] M. Peng, X. Ding, J. Qin, J. Wang, X. Bi, D. Wang, L. Luo, H. Zhao, C. Zhang, Z. Lin, L. Hong, L. Zhang, J. Chen, C. Liu, Y. Chen, Y. Cai, Q. Zhu, J. Jiang, L. Yang, S. Yu, X. Wu,

Z. Zheng, S. Fong, Q. Zhao, S. Chen, and J. Li, The Lancet (2020).

[39] X. Zhang, Q. Qi, Q. Jing, S. Ao, Z. Zhang, and M. Ding, ArXiv 1 (2020).

[40] T. Nguyen, D.D. Bang, and A. Wolff, Micromachines **0**, 1 (2020).

[41] A.F. Gavela, D.G. García, J.C. Ramirez, and L.M. Lechuga, Sensors **16**, 1 (2016).

[42] J.U. Lee, A.H. Nguyen, and J.S. Sang, Biosensors and Bioelectronic **74**, 341 (2015).

[43] M. Soler, C.S. Huertas, and L.M. Lechuga, Expert Review of Molecular Diagnostics (2019).

[44] J. Masson, Acs Sensors **2**, 16 (2017).

[45] H.H. Nguyen, J. Park, S. Kang, and M. Kim, Sensors 10481 (2015).

[46] G.K. Joshi, S. Deitz-mcelyea, M. Johnson, M. Korc, and R. Sardar, Nano Letters (2014).

[47] S. Chakravarty, W. Lai, Y. Zou, H.A. Drabkin, M. Gemmill, G.R. Simon, S.H. Chin, and R.T. Chen, Biosensors and Bioelectronic **43**, 50 (2013).

[48] H.K. Hunt and A.M. Armani, Nanoscale **2**, 1544 (2010).

[49] J. Liu, M. Jalali, S. Mahshid, and S. Wachsmann-hogiu, Analyst **145**, 364 (2020).

[50] J. Luan, A. Seth, R. Gupta, Z. Wang, P. Rathi, S. Cao, H.G. Derami, R. Tang, B. Xu, S. Achilefu, J.J. Morrissey, and S. Singamaneni, Nat Biomed Eng **4**, 518 (2020).

[51] S. Afsahi, M.B. Lerner, J.M. Goldstein, J. Lee, X. Tang, D.A. Bagarozzi, D. Pan, L. Locascio, A. Walker, F. Barron, and B.R. Goldsmith, Biosensors and Bioelectronic **100**, 85 (2018).

[52] N. Gao, T. Gao, X. Yang, X. Dai, W. Zhou, A. Zhang, and C.M. Lieber, PNAS **113**, 14633 (2016).

[53] J.A. Jackman, E. Linardy, D. Yoo, J. Seo, W. Ng, D.J. Klemme, N.J. Wittenberg, S. Oh, and N. Cho, Small 1159 (2016).

[54] M. Vestergaard, K. Kerman, and E. Tamiya, Sensors **7**, 3442 (2007).

[55] A. Leung, P.M. Shankar, and R. Mutharasan, Sensors and Actuators B **125**, 688 (2007).

[56] X. Fan, I.M. White, S.I. Shopova, H. Zhu, J.D. Suter, and Y. Sun, Analytica Chimica Acta

**0**, 8 (2008).

[57] J.G. Wangüemert-pérez, A. Hadij-elhouati, A. Sánchez-postigo, J. Leuermann, D. Xu, P. Cheben, A. Ortega-moñux, R. Halir, and I. Molina-Fernandez, Optics and Laser Technology Journal **109**, 437 (2019).

[58] H. Zhang, N. Healy, L. Shen, C.C. Huang, D.W. Hewak, and A.C. Peacock, Scientific Reports 2 (2016).

[59] W. Lai, S. Chakravarty, Y. Zou, and R.T. Chen, Optics Letters **37**, 1208 (2012).

[60] C. Yang, H. Yan, N. Tang, Y. Zou, Y. Al-hadeethi, and X. Xu, Micromachines **11**, 282 (2020).

[61] Y. Zou, S. Chakravarty, S. Member, D.N. Kwong, W. Lai, X. Xu, X. Lin, A. Hosseini, and R.T. Chen, IEEE Journal of Selected Topics in Quantum Electronics **20**, (2014).

[62] S. Chakravarty, X. Chen, N. Tang, W. Lai, Y. Zou, and H. Yan, Front. Optoelectron **9**, 206 (2016).

[63] S. Chakravarty, Y. Zou, W. Lai, and R.T. Chen, Biosensors and Bioelectronic **38**, 170 (2012).

[64] Y. Zou, S. Chakravarty, L. Zhu, and R.T. Chen, Applied Physics Letters **141103**, 1 (2014).

[65] H. Yan, Y. Zou, S. Chakravarty, C. Yang, Z. Wang, N. Tang, D. Fan, and R.T. Chen, Applied Physics Letters **106**, (2015).

[66] Y. Zou, S. Chakravarty, W. Lai, C. Lin, and R.T. Chen, Lab on a Chip 2309 (2012).

[67] S. Pal, A.R. Yadav, M.A. Lifson, J.E. Baker, P.M. Fauchet, and B.L. Miller, Biosensors and Bioelectronic **44**, 229 (2013).

[68] W. Lai, S. Chakravarty, Y. Zou, Y. Guo, R.T. Chen, W. Lai, S. Chakravarty, Y. Zou, Y. Guo, and R.T. Chen, Applied Physics Letters **102**, (2013).

[69] C. Chang, X. Xu, S. Chakravarty, H. Huang, L. Tu, Q. Yungsung, H. Dalir, M.A. Krainak, and R.T. Chen, Biosensors and Bioelectronic **141**, 111396 (2019).

[70] Kehl, Florian, D. Bischof, M. Michler, M. Keka, and R. Stanley, Photonics **2**, 124 (2015).

[71] T. Mayr, T. Abel, B. Enko, S. Borisov, C. Konrad, K. Stefan, B. Lamprecht, S. Sax, E.J.W. List, and I. Klimant, Analyst **134**, 1544 (2009).

[72] R. Yan, S.P. Mestas, G. Yuan, R. Safaisini, S. Dandy, and K.L. Lear, Lab on a Chip **9**, (2009).

[73] L. Liu, X. Zhou, J.S. Wilk, P. Hua, B. Song, and H. Shi, Scientific Reports 1 (2017).

[74] H. Lin, Z. Luo, T. Gu, L.C. Kimerling, K. Wada, A. Agarwal, and J. Hu, Nanophotonics (2017).

[75] C. Chen and J. Wang, Analyst **145**, 1605 (2020).

[76] Q. Liu, X. Tu, K. Woo Kim, J. Sheng, Y. Shin, K. Han, Y. Yoon, G. Lo, and M. Kyoung Park, Sensors & Actuators: B. Chemical **188**, 681 (2013).

[77] Q. Liu, Y. Shin, J. Sheng, K. Woo, S. Rafeah, M. Rafei, A. Promoda, X. Tu, G. Lo, E. Ricci, M. Colombel, E. Chiong, J. Paul, and M. Kyoung, Biosensors and Bioelectronic **71**, 365 (2015).

[78] T. Chalyan, R. Guider, L. Pasquardini, M. Zanetti, F. Falke, E. Schreuder, R.G. Heideman, C. Pederzolli, and L. Pavesi, Biosensors **6**, 1 (2016).

[79] H. Chen, C. Wang, H. Ouyang, Y. Song, and T. Jiang, Nanophotonics (2020).

[80] S.T. Fard, V. Donzella, S.A. Schmidt, J. Flueckiger, S.M. Grist, P. Talebifard, Y. Wu, R.J. Bojko, E. Kwok, N.A.F. Jaeger, D.M. Ratner, and L. Chrostowski, Optics Express **22**, 9499 (2014).

[81] J.A.D.E. Feijter, J. Benjamins, and F.A. Veer, Biopolymers **17**, 1759 (1978).

[82] Z. Wang, X. Xu, D. Fan, Y. Wang, and R.T. Chen, Optics Letters **41**, 3375 (2016).

[83] Z. Wang, X. Xu, D. Fan, Y. Wang, H. Subbaraman, and R.T. Chen, Scientific Reports 1 (2016).

[84] Y. Hai, L. Huang, X. Xiaochuan, S. Chakravarty, N. Tang, H. Tian, R.T. Chen, H. Yan, L. Huang, X. Xiaochuan, S. Chakravarty, N. Tang, H. Tian, and R.T. Chen, Optics Express **24**, 29724 (2016).

[85] L. Huang, H. Yan, X. Xiaochuan, S. Chakravarty, N. Tang, H. Tian, R.T. Chen, Y. Hai, X. Xu, S. Chakravarty, N. Tang, H. Tian, and R.T. Chen, Optics Express **25**, 10527 (2017).

[86] A.M. Armani, R.P. Kulkarni, S.E. Fraser, R.C. Flagan, and K.J. Vahala, Science **317**, 783 (2007).

[87] D.K. Armani, T.J. Kippenberg, S.M. Spillane, and K.J. Vahala, Nature **421**, 925 (2003).

[88] S. Chakravarty, X. Xu, H. Yan, W. Lai, Y. Zou, and R.T. Chen, Optical Sensing and Sensors 2 (2017).

[89] C. Kang, C.T. Phare, Y.A. Vlasov, S. Assefa, and S.M. Weiss, Optics Express **18**, 944 (2010).

[90] M. Lee and P.M. Fauchet, Optics Express **15**, 4530 (2007).

[91] M.G. Scullion, A. Di Falco, and T.F. Krauss, Biosensors and Bioelectronics **27**, 101 (2011).

[92] D. Dorfner, T. Zabel, T. Hürlimann, N. Hauke, L. Frandsen, U. Rant, G. Abstreiter, and J. Finley, Biosensors and Bioelectronic **24**, 3688 (2009).

[93] C.A. Barrios, Sensors **9**, 4751 (2009).

[94] K. De Vos, I. Bartolozzi, E. Schacht, P. Bienstman, and R. Baets, OPTICS EXPRESS **15**, 7610 (2007).

[95] C.F. Carlborg, K.B. Gylfason, and A. Ka, Lab on a Chip **10**, 257 (2010).

[96] A. Rahtuvanoğlu, D.S. Akgönüllü, S. Karacan, and A. Denizli, Chemistry Select **5**, 5683 (2020).

[97] A. You, M.A.Y. Be, and I. In, Applied Physics Letters **97**, 2010 (2014).

[98] M. Li, X. Wu, L. Liu, X. Fan, and L. Xu, Analytical Chemistry **85**, 9328 (2013).

[99] D.K. Nguyen and C. Jang, Analytical Biochemistry 113807 (2020).

[100] H. Sipova, S. Zhang, Aimee M, D. Galas, and K. Wang, Analytical Chemistry **82**, 10110 (2010).

[101] J.H. Jung, D.S. Cheon, F. Liu, K.B. Lee, and T.S. Seo, Angew. Chem. Int. Ed. **49**, 5708

(2010).

[102] C. Lin, X. Wang, S. Chakravarty, B.S. Lee, W. Lai, and R.T. Chen, APPLIED PHYSICS LETTERS **97**, 4 (2010).

[103] J.C. Huang, Y. Chang, K. Chen, L. Su, C. Lee, C.-C. Chen, Y.-M.A. Chen, and C. Chou, Biosensors and Bioelectronic 320 (2009).

[104] D. Rodrigo, O. Limaj, D. Janner, D. Etezadi, F.J.G. De Abajo, V. Pruneri, and H. Altug, **165**, 1 (2015).

[105] L. Shi, Q. Sun, H. Xu, C. Liu, C. Zhao, Y. Xu, C. Wu, J. Xiang, D. Gu, J. Long, and H. Lan, Bio-Medical Materials and Engineering **26**, 2207 (2015).

[106] D. Rodrigo, O. Limaj, D. Janner, D. Etezadi, F.J.G. De Abajo, V. Pruneri, and H. Altug, Science **349**, 165 (2015).

[107] G. Qiu, S.P. Ng, and L.C. Wu, Sensors and Actuators B: Chemical **234**, 247 (2016).

[108] A. Rapisarda, N. Giamblanco, and G. Marletta, Journal of Colloid And Interface Science **487**, 141 (2017).

[109] A.J. Haes, L. Chang, W.L. Klein, and R.P. Van Duyne, J. AM. CHEM. SOC. 2264 (2005).

[110] M.S. Bin-alam, O. Reshef, Y. Mamchur, M.Z. Alam, G. Carlow, J. Upham, B.T. Sullivan, M. Jean-michel, M.J. Huttunen, R.W. Boyd, and K. Dolgaleva, ArXiv 41 (2020).

[111] L. Wang, Q. Song, Q. Liu, D. He, and J. Ouyang, Advanced Functional Materials **25**, 7017 (2015).

[112] H. Ilkhani, T. Hughes, J. Li, C. Jian, and M. Hepel, Biosensors and Bioelectronic **80**, 257 (2016).

[113] L.A. Layqah and S. Eissa, Microchimica Acta **186**, 224 (2019).

[114] B. Huang, F. Yu, and R.N. Zare, Anal. Chem. **79**, 2979 (2007).

[115] Y. Tang and X. Zeng, Chemical Education **87**, 742 (2010).

[116] A. Abbas, M.J. Linman, and Q. Cheng, Biosensors and Bioelectronics **26**, 1815 (2011).

[117] J. Homola, Anal Bioanal Chem **377**, 528 (2003).

[118] J. Homola, Chem. Rev. **108**, 462 (2008).

[119] B. Sciacca, A. François, M. Klingler-hoffmann, J. Brazzatti, M. Penno, P. Hoffmann, and T.M. Monro, Nanomedicine: Nanotechnology, Biology, and Medicine **9**, 550 (2013).

[120] J.T. Hastings, **8**, 170 (2008).

[121] S. Akter, Z. Rahman, and S. Mahmud, Results in Physics **13**, 102328 (2019).

[122] H. Heidarzadeh, Optics Communications **459**, 124940 (2020).

[123] S.S. Acimovic, M.A. Ortega, V. Sanz, J. Berthelot, J.L. Garcia-cordero, J. Renger, S.J. Maerkl, M.P. Kreuzer, and R. Quidant, Nano Letters **14**, 2636 (2014).

[124] P. Chen, M.T. Chung, W. Mchugh, R. Nidetz, Y. Li, J. Fu, T.T. Cornell, T.P. Shanley, and K. Kurabayashi, ACS Nano 4173 (2015).

[125] H. Yu, Y. Peng, Y. Yang, and Z. Li, Npj Computational Materials 1 (2019).

[126] A. Djaileb, B. Charron, and M. Hojjat, ChemRxiv 1 (2020).

[127] T. Yoshie, L. Tang, and S. Su, Sensors **11**, 1972 (2011).

[128] D.J. Bergman and M.I. Stockman, PHYSICAL REVIEW LETTERS **90**, 1 (2003).

[129] A.A. Kolomenskii, P.D. Gershon, and H.A. Schuessler, **36**, 6539 (1997).

[130] N.I. Landy, S. Sajuyigbe, J.J. Mock, D.R. Smith, and W.J. Padilla, PHYSICAL REVIEW LETTERS **100**, 207402(1 (2008).

[131] H. Tao, C.M. Bingham, A.C. Strikwerda, D. Pilon, D. Shrekenhamer, N.I. Landy, K. Fan, X. Zhang, W.J. Padilla, and R.D. Averitt, Physical Review B **78**, 2 (2008).

[132] A.A. Jamali and B. Witzigmann, Plasmonics 1265 (2014).

[133] N. Liu, M. Mesch, T. Weiss, M. Hentschel, and H. Giessen, Nano Letters **10**, 2342 (2010).

[134] S. Korkmaz, M. Turkmen, and S. Aksu, Sensors & Actuators: A. Physical **301**, 111757 (2020).

[135] V. Giannini, Y. Francescato, H. Amrania, C.C. Phillips, and S.A. Maier, Nano Letters **11**, 2835 (2011).

[136] D. Yoo, D.A. Mohr, F. Vidal-codina, A. John-herpin, M. Jo, S. Kim, J. Matson, J.D. Caldwell, H. Jeon, N. Nguyen, L. Martin-moreno, J. Peraire, H. Altug, and S. Oh, Nano Letters (2018).

[137] X. Chen, H. Park, M. Pelton, X. Piao, N.C. Lindquist, H. Im, Y.J. Kim, J.S. Ahn, K.J. Ahn, N. Park, D. Kim, and S. Oh, Nature Communications 1 (2013).

[138] D. Yoo, N. Nguyen, L. Martin-moreno, D.A. Mohr, S. Carretero-palacios, J. Shaver, J. Peraire, T.W. Ebbesen, and S. Oh, (2016).

[139] H. Park, X. Chen, N. Nguyen, J. Peraire, and S. Oh, Acs Photonics **2**, 417 (2015).

[140] Y. Shen, J. Zhou, T. Liu, Y. Tao, R. Jiang, M. Liu, G. Xiao, J. Zhu, Z. Zhou, X. Wang, C. Jin, and J. Wang, Nature Communications (2013).

[141] E. Suenaga, H. Mizuno, and P.K.R. Kumar, Virulence **3**, 464 (2012).

[142] S. Zeng, K.V. Sreekanth, J. Shang, T. Yu, C. Chen, F. Yin, D. Baillargeat, P. Coquet, H. Ho, A. V Kabashin, and K. Yong, Advanced Materials **27**, 6163 (2015).

[143] G.Y. Jia, Z.X. Huang, Y.L. Zhang, Z.Q. Hao, and Y.L. Tian, J. Mater. Chem. C **7**, 3843 (2019).

[144] S. Kim and H.J. Lee, (2017).

[145]News 18 (2020).

[146] K.S. Novoselov, A.K. Geim, and S. V Morozov, Science **22**, 2 (2004).

[147] I. Zand, H. Dalir, R.T. Chen, and J.P. Dowling, Applied Physics Express **11**, 035101 (1 (2018).

[148] E. Heidari, Z. Ma, H. Dalir, V. Esfandyarpour, V.J. Sorger, and R.T. Chen, in *Proc. SPIE 10924, Optical Interconnects XIX, 1092419* (2019).

[149] F. Bonaccorso, Z. Sun, T. Hasan, and A.C. Ferrari, Nature Photonics **4**, 611 (2010).

[150] N.O. Weiss, H. Zhou, L. Liao, Y. Liu, S. Jiang, Y. Huang, and X. Duan, Advanced Materials **24**, 5782 (2012).

[151] Y. Pang, J. Jian, T. Tu, Z. Yang, J. Ling, Y. Li, X. Wang, Y. Qiao, H. Tian, Y. Yang, and

T. Ren, Biosensors and Bioelectronic **116**, 123 (2018).

[152] J. Peña-Bahamonde, H.N. Nguyen, S.K. Fanourakis, and D.F. Rodrigues, Journal of Nanobiotechnology 1 (2018).

[153] A. Asghari, H. Dalir, Sorger, Volker, and R.T. Chen, in *International Society for Optics and Photonics* (2020).

[154] Z. Ma, K. Kikunaga, H. Wang, S. Sun, R. Amin, R. Maiti, M.H. Tahersima, H. Dalir, M. Miscuglio, and V.J. Sorger, Acs Photonics **7**, 932 (2020).

[155] P.M. Grubb, F.M. Koushyar, T. Lenz, A. Asghari, G. Gan, W. Xia, H. Dalir, H. Subbaraman, and R.T. Chen, (n.d.).

[156] H. Dalir, Y. Xia, Y. Wang, and X. Zhang, Acs Photonics **3**, 1564 (2016).

[157] A. Asghari, H. Dalir, V. Sorger, and R.T. Chen, in *2D Photonic Materials and Devices* (2020).

[158] H. Dalir, E. Heidari, A. Asghari, M.H. Teimourpour, V.J. Sorger, and R.T. Chen, in *Society for Optics and Photonics.* (International Society for Optics and Photonics, 2020).

[159] R. Amin, S. Khan, C.J. Lee, H. Dalir, and V.J. Sorger, 5 (2018).

[160] V.J. Sorger, R. Maiti, M.H. Tahersima, R. Hemnani, H. Dalir, and R. Agarwal, in *Proc. SPIE 10920, 2D Photonic Materials and Devices II, 109200H (8 March 2019)* (2019).

[161] Z. Ma, K. Kikunaga, H. Wang, S. Sun, R. Amin, M. Miscuglio, H. Dalir, and V.J. Sorger, in *Proc. SPIE 10927, Photonic and Phononic Properties of Engineered Nanostructures IX, 109270V* (2019).

[162] N. Chauhan, T. Maekawa, D. Nair, and S. Kumar, J. Mater. Res. **32**, 2860 (2017).

[163] S. Zhang, Z. Li, and F. Xing, International Journal of Molecular Sciences Review **2**, (2020).

[164] L. Wu, H.S. Chu, W.S. Koh, and E.P. Li, Optics Express **18**, 15458 (2010).

[165] P.K. Ang, A. Li, M. Jaiswal, Y. Wang, H.W. Hou, J.T.L. Thong, C.T. Lim, and K.P. Loh, Nano Letters **11**, 5240 (2011).

[166] H. Huang, S. Su, N. Wu, H. Wan, S. Wan, H. Bi, and L. Sun, Frontiers in Chemistry **7**, 1 (2019).

[167] G. Nikoleli, D.P. Nikolelis, C. Siontorou, S. Karapetis, S. Bratakou, and N. Tzamtzis, *Nanobiosensors Based on Graphene Electrodes : Recent Trends and Future Applications* (Elsevier Ltd., 2018).

[168] J. Kampeera, P. Pasakon, C. Karuwan, and N. Arunrut, Biosensors and Bioelectronic **132**, 271 (2019).

[169] S. Jin, S. Poudyal, E.E. Marinero, R.J. Kuhn, and L.A. Stanciu, Electrochimica Acta **194**, 422 (2016).

[170] F. Xing, Z. Liu, Z. Deng, X. Kong, X. Yan, X. Chen, Q. Ye, C.-P. Zhang, Y.-S. Chen, and J.-G. Tian, Scientific Reports 1 (2012).

[171] M. Liu, X. Yin, E. Ulin-avila, B. Geng, T. Zentgraf, L. Ju, F. Wang, and X. Zhang, Nature **474**, (2011).

[172] J. Wang, Z. Xing, X. Chen, Z. Cheng, X. Li, and T. Liu, Frontiers in Physics **8**, 1 (2020).

[173] I.A. Calafell, J.D.C. M. Radonjić, J.R.M. Saavedra, F.J. Garcia de Abajo, L.A. Rozema, and P. Walther, Npj Quantum Information **5**, 37 (2019).

[174] K.J.A. Ooi and D.T.H. Tan, Proc. R. Soc. **473**, (2017).

[175] G. Saltzgaber, P.M. Wojcik, T. Sharf, M.R. Leyden, J.L. Wardini, C.A. Heist, A.A. Adenuga, V.T. Remcho, and E.D. Minot, Nanotechnology **24**, (2013).

[176] J.E. Lee, G. Ahn, J. Shim, Y.S. Lee, and S. Ryu, Nature Communications **3**, 1024 (2012).

[177] H. He, K.H. Kim, A. Danilov, D. Montemurro, L. Yu, Y.W. Park, F. Lombardi, T. Bauch, K. Moth-poulsen, T. Iakimov, R. Yakimova, P. Malmberg, C. Müller, S. Kubatkin, and S. Lara-avila, Nature Communications **9**, 3 (2018).

[178] B. Liu, C. Yang, Z. Liu, and C. Lai, Nanomaterials **7**, 302 (2017).

[179] A. Asghari, N. Antanio, U. Castillo, E. Segura-cardenas, and A.O. Sustaita, Materials Research Express **6**, 0 (2019).

[180] A. Bolotsky, D. Butler, C. Dong, K. Gerace, and N.R. Glavin, (2019).